\documentclass[aps,prd,reprint,floatfix,amsmath,nofootinbib,amssymb,superscriptaddress]{revtex4-2}

\usepackage{graphicx}
\usepackage{xcolor}

\usepackage{bm}
\usepackage{slashed}
\allowdisplaybreaks[4]

\usepackage{booktabs}
\usepackage{multirow}

\usepackage[
  unicode=true,
  bookmarks=true,
  bookmarksopen=true,
  bookmarksnumbered=true,
  colorlinks=true,
  linkcolor=blue,
  citecolor=blue,
  urlcolor=blue
]{hyperref}
\usepackage{bookmark}

\makeatletter
\renewcommand*{\addcontents@abstract}{}
\makeatother

\begin{document}

\title{Probing tensor-polarized gluon TMDs through \texorpdfstring{$J/\psi$}{J/psi} electroproduction off the deuteron}

\author{Xiupeng Xie}
\email{xiexiupeng@seu.edu.cn}
\affiliation{School of Physics, Southeast University, Nanjing 211189, China}

\author{Zhun Lu}
\email{zhunlu@seu.edu.cn}
\affiliation{School of Physics, Southeast University, Nanjing 211189, China}

\begin{abstract}
We investigate tensor-polarized gluon transverse-momentum-dependent distributions (TMDs) in semi-inclusive $J/\psi$ electroproduction off the deuteron. At leading order in the color-octet channels of nonrelativistic QCD, we derive the cross-section contributions associated with longitudinal-longitudinal ($LL$), longitudinal-transverse ($LT$), and transverse-transverse ($TT$) tensor polarization and define ten normalized observables for unpolarized and longitudinally polarized electron beams. Using gluon TMDs obtained in a spin-1 spectator model, we study their $q_T$, $y$, $Q^2$, and collision-energy dependences. The observables involving $f_{1LL}^g$, $f_{1LT}^g+h_{1LT}^g$, and $f_{1TT}^g-h_{1TT}^{\perp g}$ depend directly on ratios of gluon TMDs. The moments involving linearly polarized gluons exhibit a common zero at $Q^2=M_\psi^2$ for the adopted long-distance matrix elements, whereas the beam-helicity-dependent sine moments probe the T-odd distributions $g_{1LT}^g$ and $g_{1TT}^g$. We present fixed-scale model estimates that neglect transverse recoil from soft radiation and quarkonium formation.
\end{abstract}

\maketitle

\section{Introduction}
\label{sec:introduction}

The deuteron provides access to partonic spin correlations associated with the tensor polarization of a spin-1 hadron. Its spin density matrix contains a symmetric traceless tensor in addition to vector polarization, thereby allowing tensor-polarized parton distributions to be defined~\cite{Hoodbhoy:1988am,Bacchetta:2000jk}. The deuteron structure function $b_1$ and the corresponding quark distributions have been studied through sum rules~\cite{Close:1990zw}, experimental measurements~\cite{HERMES:2005pon}, phenomenological extractions, and convolution calculations~\cite{Kumano:2010vz,Cosyn:2017fbo}. The gluon sector contains additional correlations, including the tensor-polarized double-helicity-flip distribution~\cite{Jaffe:1989xy}. Proposed proton-deuteron Drell-Yan measurements offer a complementary probe of this distribution~\cite{Kumano:2019igu,Kumano:2020gfk}.

Gluon transverse-momentum-dependent distributions (TMDs) retain information on the partonic transverse motion and its correlations with spin~\cite{Mulders:2000sh,Meissner:2007rx}. Their parametrization extends to tensor-polarized spin-1 hadrons~\cite{Boer:2016xqr,Cotogno:2017puy}, while gauge links encode process dependence arising from initial- and final-state interactions~\cite{Collins:2002kn,Buffing:2013kca}. Heavy-quark pair and dijet production provide access to gluon TMDs through azimuthal modulations~\cite{Boer:2010zf,Pisano:2013cya,Boer:2016fqd}, and quarkonium transverse-momentum spectra offer complementary sensitivity~\cite{Boer:2012bt}. In semi-inclusive $J/\psi$ electroproduction, photon-gluon fusion connects the angular distribution to gluon TMDs when the quarkonium transverse momentum is small compared with the hard scales~\cite{Mukherjee:2016qxa,Bacchetta:2018ivt}.

Within nonrelativistic QCD (NRQCD), heavy-quark-pair production is described by short-distance coefficients, whereas the transition into quarkonium is encoded in long-distance matrix elements~\cite{Bodwin:1994jh}. At the lowest order for $\gamma^*g\to c\bar c$, the relevant color-octet channels are ${}^{1}S_0^{[8]}$ and ${}^{3}P_J^{[8]}$, with $J=0,1,2$. Studies of $J/\psi$ electroproduction have addressed the gluon Sivers and $\cos2\phi_T$ asymmetries~\cite{Mukherjee:2016qxa,Bacchetta:2018ivt,Kishore:2018ugo}, associated $J/\psi+$jet production~\cite{DAlesio:2019qpk}, and TMD evolution~\cite{Bor:2022fga}. Longitudinally polarized scattering also provides access to $h_{1L}^{\perp g}$ and $g_{1L}^{g}$~\cite{liu:2023unl}. The vector-polarized gluon responses provide a check of the hard-scattering calculation and include a beam-helicity-dependent contribution from $g_{1T}^g$. A tensor-polarized deuteron introduces additional longitudinal-longitudinal ($LL$), longitudinal-transverse ($LT$), and transverse-transverse ($TT$) responses, whose relation to polarization observables requires a dedicated analysis.

Spectator models have provided estimates of T-even and T-odd gluon TMDs in the nucleon~\cite{Lu:2016vqu,Bacchetta:2020vty,Bacchetta:2024fci}. Related spectator-model studies have also addressed twist-3 gluon TMDs in the nucleon and both T-even and T-odd gluon TMD fragmentation functions~\cite{Xie:2026dmo,Xie:2022lra,Xie:2024cnv}. In our previous work, we calculated the leading-twist T-even gluon TMDs of the deuteron in a spin-1 spectator model~\cite{Xie:2026uja}. The model employs a continuous spectator-mass spectrum and parameters constrained by the unpolarized deuteron gluon distribution. A subsequent calculation included the final-state interaction needed to generate T-odd gluon TMDs~\cite{Xie:2026zjn}. Together, these results provide nonperturbative input for estimating the size and momentum dependence of the tensor-polarized contributions to $J/\psi$ electroproduction.

Here we consider $e+d\to e'+J/\psi+X$ with a polarized deuteron and an unpolarized or longitudinally polarized electron beam. We present the unpolarized and vector-polarized cross sections alongside the $LL$, $LT$, and $TT$ tensor-polarized contributions. The former reproduce the established structures for spin-1/2 targets and retain the nonzero beam-helicity-dependent $g_{1T}^g$ term. We define normalized tensor-polarization observables from suitable polarization combinations and angular weights. The normalization is provided by the unpolarized cross section, with the beam and target polarization factors removed. Using our spectator-model gluon TMDs, we examine the $q_T$, $y$, $Q^2$, and $\sqrt{s_{eD}}$ dependences, focusing in particular on the contributions of $f_{1LL}^g$, $h_{1LT}^g$, $h_{1TT}^g$, and $g_{1TT}^g$. The results may be relevant for studies of deuteron spin structure at future electron-ion colliders~\cite{Accardi:2012qut,AbdulKhalek:2021gbh,Anderle:2021wcy}.

Our calculation uses leading-order photon-gluon fusion and neglects transverse momentum generated by soft radiation and quarkonium formation. The effects of soft radiation and their treatment in quarkonium production have been studied in Refs.~\cite{Boer:2020bbd,Echevarria:2019ynx,Fleming:2019pzj}. A more complete treatment involves quarkonium TMD shape functions~\cite{Boer:2023zit,Echevarria:2024idp,Echevarria:2025oab}. The present results should therefore be regarded as model estimates within this approximation.

The electroproduction formalism and observable definitions are presented in Sec.~\ref{sec:formalism}. Section~\ref{sec:setup} specifies the model input and numerical settings. The results and their interpretation are discussed in Sec.~\ref{sec:results}, followed by the conclusions in Sec.~\ref{sec:summary}.

\section{Polarization-dependent \texorpdfstring{$J/\psi$}{J/psi} electroproduction}
\label{sec:formalism}

We consider the semi-inclusive process
\begin{align}
e(\ell,\lambda_e)+d(P,S,T)\to e(\ell')+J/\psi(P_\psi)+X\,.
\label{eq:process}
\end{align}
The quarkonium polarization is summed over. The deuteron mass is denoted by $M_D$, whereas $M_\psi$ denotes the quarkonium mass. The spin vector $S$ and the symmetric traceless spin tensor $T$ describe vector and tensor polarization, respectively. We give the unpolarized, vector-polarized ($L$ and $T$), and tensor-polarized ($LL$, $LT$, and $TT$) contributions to the cross section. The numerical analysis below focuses on tensor-polarized observables.

The standard variables are
\begin{align}
q&=\ell-\ell',& Q^2&=-q^2,\notag\\
x_B&=\frac{Q^2}{2P\cdot q},&y&=\frac{P\cdot q}{P\cdot\ell},\notag\\
z&=\frac{P\cdot P_\psi}{P\cdot q},&s_{eD}&=(P+\ell)^2\,.
\label{eq:kinematics}
\end{align}
In a frame where $P$ and $q$ define the longitudinal plane, we write $\bm q_T=\bm P_{\psi T}$ and measure its azimuth $\phi_T$ from the lepton plane. The transverse $x$ direction follows the incoming lepton transverse momentum. We use $g_T^{ij}=-\delta^{ij}$ and $\epsilon_T^{12}=+1$. The transverse vector polarization and the tensor-polarization directions are specified by
\begin{align}
\bm S_T&=s_T(\cos\phi_S,\sin\phi_S)\,,\notag\\
\bm S_{LT}&=s_{LT}(\cos\phi_{LT},\sin\phi_{LT})\,,\notag\\
S_{TT}^{ij}&=s_{TT}
\begin{pmatrix}
\cos2\phi_{TT}&\sin2\phi_{TT}\\
\sin2\phi_{TT}&-\cos2\phi_{TT}
\end{pmatrix},
\label{eq:tensor-angles}
\end{align}
where $s_T=|\bm S_T|$, $s_{LT}=|\bm S_{LT}|$, and $s_{TT}=\sqrt{(S_{TT}^{11})^2+(S_{TT}^{12})^2}$. The $TT$ orientation is defined modulo $\pi$. Our normalization of $S_{LL}$ gives $S_{LL}=1/2$ for $m=\pm1$ and $S_{LL}=-1$ for $m=0$ when the spin is quantized along the longitudinal axis~\cite{Bacchetta:2000jk,Boer:2016xqr}. Tensor polarization is even under reversal of the spin-quantization axis, so an ordinary vector-spin flip does not by itself isolate these responses. General tensor-polarized SIDIS angular decompositions are discussed in Ref.~\cite{Zhao:2025vol}.

At the order considered here, the hard subprocess is $\gamma^*g\to c\bar c[n]$, followed by the transition into $J/\psi$. In the approximation of vanishing transverse recoil during this transition,
\begin{align}
z&=1,\qquad \bm k_T=\bm q_T,\notag\\
x_g&=x_B\left(1+\frac{M_\psi^2}{Q^2}\right)
\simeq\frac{Q^2+M_\psi^2}{y s_{eD}}\,.
\label{eq:xg}
\end{align}
Here $x_g$ is the gluon momentum fraction relative to the entire deuteron. It is this variable that enters the spectator-model distributions. We neglect target-mass and transverse-momentum power corrections to Eq.~\eqref{eq:xg}, with the leading-power expressions intended to apply for $q_T\ll Q,M_\psi$. The differential cross sections presented below are obtained by integrating over $z$ the leading-order contribution proportional to $\delta(1-z)$. They describe the small-$q_T$ color-octet contribution rather than the full measured $z$ spectrum~\cite{Bacchetta:2018ivt,Boer:2020bbd,Echevarria:2024idp}.

We use the leading-twist gluon TMD parametrization of the spectator-model calculation~\cite{Xie:2026uja,Xie:2026zjn}. With the overall convention $\Gamma^{ij}=x_g\Phi^{ij}$, the six polarization components read
\begin{align}
\Gamma_U^{ij}&=\frac{x_g}{2}\left[-g_T^{ij}f_1^g
+\frac{k_T^{ij}}{M_D^2}h_1^{\perp g}\right]\,,\label{eq:corr-u}\\
\Gamma_L^{ij}&=\frac{x_g}{2}S_L\Bigg[i\epsilon_T^{ij}g_{1L}^g+\frac{\epsilon_{T\alpha}^{\{i}k_T^{j\}\alpha}}{2M_D^2}h_{1L}^{\perp g}\Bigg]\,,\label{eq:corr-l}\\
\Gamma_T^{ij}&=\frac{x_g}{2}\Bigg[-g_T^{ij}\frac{\epsilon_T^{S_Tk_T}}{M_D}
f_{1T}^{\perp g}+i\epsilon_T^{ij}\frac{\bm k_T\cdot\bm S_T}{M_D}g_{1T}^g\notag\\
&\quad-\frac{\epsilon_T^{k_T\{i}S_T^{j\}}+\epsilon_T^{S_T\{i}k_T^{j\}}}{4M_D}h_1^g\notag\\
&\quad-\frac{\epsilon_{T\alpha}^{\{i}k_T^{j\}\alpha\beta}S_{T\beta}}{2M_D^3}h_{1T}^{\perp g}\Bigg]\,,\label{eq:corr-t}\\
\Gamma_{LL}^{ij}&=\frac{x_g}{2}S_{LL}\left[-g_T^{ij}f_{1LL}^g
+\frac{k_T^{ij}}{M_D^2}h_{1LL}^{\perp g}\right]\,,\label{eq:corr-ll}\\
\Gamma_{LT}^{ij}&=\frac{x_g}{2}\Bigg[-\frac{g_T^{ij}\bm k_T\cdot\bm S_{LT}}{M_D}f_{1LT}^g\notag\\
&\quad+\frac{i\epsilon_T^{ij}\epsilon_T^{S_{LT}k_T}}{M_D}g_{1LT}^g
+\frac{S_{LT}^{\{i}k_T^{j\}}}{M_D}h_{1LT}^g\notag\\
&\quad+\frac{k_T^{ij\alpha}S_{LT\alpha}}{M_D^3}h_{1LT}^{\perp g}\Bigg]\,,\label{eq:corr-lt}\\
\Gamma_{TT}^{ij}&=\frac{x_g}{2}\Bigg[-\frac{g_T^{ij}k_T^{\alpha\beta}S_{TT\alpha\beta}}{M_D^2}f_{1TT}^g\notag\\
&\quad+\frac{i\epsilon_T^{ij}\epsilon_{T\gamma}^{\ \ \beta}k_T^{\gamma\alpha}S_{TT\alpha\beta}}{M_D^2}g_{1TT}^g
+S_{TT}^{ij}h_{1TT}^g\notag\\
&\quad+\frac{S_{TT\alpha}^{\{i}k_T^{j\}\alpha}}{M_D^2}h_{1TT}^{\perp g}
+\frac{k_T^{ij\alpha\beta}S_{TT\alpha\beta}}{M_D^4}h_{1TT}^{\perp\perp g}\Bigg]\,.
\label{eq:corr-tt}
\end{align}
The arguments $(x_g,\bm k_T^2)$ and the gauge links are suppressed. Curly brackets denote a sum over the two index permutations without a factor $1/2$, and $k_T^{i_1\cdots i_n}$ are symmetric traceless transverse tensors. In particular, $k_T^{ij}=k_T^ik_T^j+\bm k_T^2g_T^{ij}/2$. The abbreviation $\epsilon_T^{ab}=\epsilon_T^{ij}a_i b_j$ fixes the transverse orientation. The $h_{1LT}^g$ term in Eq.~\eqref{eq:corr-lt} is written without subtracting its transverse trace. Consequently, the trace response contains $f_{1LT}^g+h_{1LT}^g$. Likewise, Eq.~\eqref{eq:corr-tt} yields the combination $f_{1TT}^g-h_{1TT}^{\perp g}$. These combinations are tied to the displayed basis; replacing it with another parametrization requires a corresponding redefinition of the scalar functions.

For this color-octet electroproduction channel, the gluon distributions carry two future-pointing gauge links, conventionally denoted $[+,+]$~\cite{Bacchetta:2018ivt,Buffing:2013kca}. The T-odd distributions must be used with this link structure and with the transverse convention in Eqs.~\eqref{eq:tensor-angles}--\eqref{eq:corr-tt}. The functions $g_{1LT}^g$ and $g_{1TT}^g$ are T-odd; the other tensor distributions entering the following unpolarized-beam expressions are T-even.

For a longitudinally polarized electron beam, the leptonic tensor can be decomposed into symmetric and antisymmetric parts,
\begin{align}
L^{\mu\nu}
=&e^2\Big[-g^{\mu\nu}Q^2
+2(\ell^\mu\ell'^\nu+\ell^\nu\ell'^\mu)
\notag\\
&+2i\lambda_e\epsilon^{\mu\nu\rho\sigma}
\ell_\rho\ell'_\sigma\Big]\,.
\label{eq:leptonic-tensor}
\end{align}
Here $\lambda_e$ denotes the longitudinal electron-beam polarization. The antisymmetric part gives rise to the beam-helicity-dependent cross-section contributions.

The differential cross section is obtained by contracting the leptonic tensor and the gluon correlator with the corresponding NRQCD hard-scattering amplitudes, following Refs.~\cite{Bacchetta:2018ivt,liu:2023unl}.

The short-distance contributions involve the states ${}^1S_0^{[8]}$ and ${}^3P_J^{[8]}$. We define
\begin{align}
\mathcal O_S&=\langle0|\mathcal O_8^{J/\psi}({}^1S_0)|0\rangle,\notag\\
\mathcal O_P&=\langle0|\mathcal O_8^{J/\psi}({}^3P_0)|0\rangle,
\label{eq:ldme-def}
\end{align}
and apply the leading heavy-quark spin-symmetry relation $\langle\mathcal O_8^{J/\psi}({}^3P_J)\rangle=(2J+1)\mathcal O_P$~\cite{Bodwin:1994jh}. The four hard functions then read
\begin{align}
\mathcal A_1&=\mathcal O_S+\frac{4(7M_\psi^2+3Q^2)}{M_\psi^2(M_\psi^2+Q^2)}\mathcal O_P,\label{eq:a1}\\
\mathcal A_2&=\frac{32Q^2}{(M_\psi^2+Q^2)^2}\mathcal O_P,\label{eq:a2}\\
\mathcal B&=-\mathcal O_S+\frac{4(3M_\psi^2-Q^2)}{M_\psi^2(M_\psi^2+Q^2)}\mathcal O_P,\label{eq:b}\\
\mathcal C&=\mathcal O_S+\frac{4(3Q^2-M_\psi^2)}{M_\psi^2(M_\psi^2+Q^2)}\mathcal O_P.
\label{eq:c}
\end{align}
These functions follow from the same photon-gluon hard scattering used for unpolarized and vector-polarized targets~\cite{Bacchetta:2018ivt,liu:2023unl}. Tensor polarization changes the gluon correlator and the angular structures, while the short-distance functions remain common. For compactness, we introduce
\begin{align}
A_0&=[1+(1-y)^2]\mathcal A_1-y^2\mathcal A_2,\notag\\
B_0&=(1-y)\mathcal B,\notag\\
C_0&=[(1-y)^2-1]\mathcal C,\notag\\
\mathcal N&=\frac{(2\pi)^2\alpha^2\alpha_s e_c^2}
{yQ^2M_\psi(M_\psi^2+Q^2)}\,,
\label{eq:hard-short}
\end{align}
where $e_c=2/3$.

For brevity, we write $\mathrm{d}\sigma$ for the differential cross section $\mathrm{d}\sigma/(\mathrm{d}x_B\,\mathrm{d}y\,\mathrm{d}^2\bm q_T)$, and use the same differential measure for every term below. The polarization decomposition of the cross section is
\begin{align}
\mathrm{d}\sigma={}&\mathrm{d}\sigma_{UU}
+S_L\big(\mathrm{d}\sigma_{U,L}+\lambda_e\mathrm{d}\sigma_{L,L}\big)\notag\\
&+s_T\big(\mathrm{d}\sigma_{U,T}+\lambda_e\mathrm{d}\sigma_{L,T}\big)\notag\\
&+S_{LL}\mathrm{d}\sigma_{U,LL}\notag\\
&+s_{LT}\big(\mathrm{d}\sigma_{U,LT}+\lambda_e\mathrm{d}\sigma_{L,LT}\big)\notag\\
&+s_{TT}\big(\mathrm{d}\sigma_{U,TT}+\lambda_e\mathrm{d}\sigma_{L,TT}\big)\,.
\label{eq:master}
\end{align}
The first subscript in $\mathrm{d}\sigma_{a,i}$ specifies the electron-beam polarization ($a=U,L$), whereas the second specifies the deuteron-polarization component ($i=U,L,T,LL,LT,TT$). The factors $S_L$, $s_T$, $S_{LL}$, $s_{LT}$, $s_{TT}$, and, where applicable, $\lambda_e$ have been extracted as in Eq.~\eqref{eq:master}. All TMDs below are evaluated at $(x_g,q_T^2)$. For an unpolarized beam and target, the differential cross section is
\begin{align}
\mathrm{d}\sigma_{UU}=\mathcal N\left[A_0 f_1^g
+\frac{q_T^2}{M_D^2}B_0 h_1^{\perp g}\cos2\phi_T\right].
\label{eq:uu}
\end{align}
The longitudinal vector-polarization responses read
\begin{align}
\mathrm{d}\sigma_{U,L}={}&\mathcal N\frac{q_T^2}{M_D^2}B_0
h_{1L}^{\perp g}\sin2\phi_T\,,\label{eq:vec-l-u}\\
\mathrm{d}\sigma_{L,L}={}&-\mathcal N C_0 g_{1L}^g\,.
\label{eq:vec-l-l}
\end{align}
For transverse vector polarization, we obtain
\begin{align}
\mathrm{d}\sigma_{U,T}=\mathcal N\Bigg[&\frac{q_T}{M_D}A_0 f_{1T}^{\perp g}
\sin(\phi_S-\phi_T)\notag\\
&+\frac{q_T}{M_D}B_0h_1^g\sin(\phi_S+\phi_T)\notag\\
&-\frac{q_T^3}{2M_D^3}B_0h_{1T}^{\perp g}
\sin(\phi_S-3\phi_T)\Bigg]\,,\label{eq:vec-t-u}\\
\mathrm{d}\sigma_{L,T}={}&-\mathcal N\frac{q_T}{M_D}C_0g_{1T}^g
\cos(\phi_S-\phi_T)\,.
\label{eq:vec-t-l}
\end{align}
Equations~\eqref{eq:uu}--\eqref{eq:vec-t-l} reproduce the unpolarized and vector-polarized angular structures of photon-gluon fusion~\cite{Bacchetta:2018ivt,liu:2023unl}, with deuteron TMDs and $M_D$ replacing their nucleon counterparts. In particular, the beam-helicity-dependent $g_{1T}^g$ term is nonzero at this order. It requires both a longitudinally polarized electron beam and a transversely vector-polarized target, and it is absent when the electron helicities are averaged.

The $LL$ response follows directly from replacing the two unpolarized-target distributions with their tensor-polarized counterparts,
\begin{align}
\mathrm{d}\sigma_{U,LL}=\mathcal N\left[A_0 f_{1LL}^g
+\frac{q_T^2}{M_D^2}B_0 h_{1LL}^{\perp g}\cos2\phi_T\right].
\label{eq:ll}
\end{align}
There is no beam-helicity-dependent $LL$ term at this order. The $LT$ responses are
\begin{align}
\mathrm{d}\sigma_{U,LT}=\mathcal N\Bigg[&\frac{q_T}{M_D}A_0(f_{1LT}^g+h_{1LT}^g)
\cos(\phi_T-\phi_{LT})\notag\\
&+\frac{2q_T}{M_D}B_0h_{1LT}^g\cos(\phi_T+\phi_{LT})\notag\\
&-\frac{q_T^3}{2M_D^3}B_0h_{1LT}^{\perp g}
\cos(3\phi_T-\phi_{LT})\Bigg],\label{eq:lt-u}\\
\mathrm{d}\sigma_{L,LT}=\mathcal N&\frac{q_T}{M_D}C_0g_{1LT}^g
\sin(\phi_T-\phi_{LT}).
\label{eq:lt-l}
\end{align}
Similarly, the $TT$ responses read
\begin{align}
\mathrm{d}\sigma_{U,TT}=\mathcal N\Bigg[&\frac{q_T^2}{M_D^2}A_0(f_{1TT}^g-h_{1TT}^{\perp g})
\cos2(\phi_T-\phi_{TT})\notag\\
&+2B_0h_{1TT}^g\cos2\phi_{TT}\notag\\
&+\frac{q_T^4}{2M_D^4}B_0h_{1TT}^{\perp\perp g}
\cos(4\phi_T-2\phi_{TT})\Bigg],\label{eq:tt-u}\\
\mathrm{d}\sigma_{L,TT}=\mathcal N&\frac{q_T^2}{M_D^2}C_0g_{1TT}^g
\sin2(\phi_T-\phi_{TT}).
\label{eq:tt-l}
\end{align}
The rank-zero $h_{1TT}^g$ term depends on the orientation of the target tensor even after integration over $\phi_T$. It is therefore accessible through a target-orientation measurement. In contrast, the higher-rank terms retain the quarkonium azimuth.

The tensor responses in Eq.~\eqref{eq:master} are coefficients of independent target-polarization components and can be isolated from the vector-polarized terms by suitable target-state combinations. For example, if $\mathrm{d}\sigma_m$ denotes the cross section for spin projection $m$ along the longitudinal axis, then
\begin{align}
\mathrm{d}\sigma_{UU}&=\frac{\mathrm{d}\sigma_{+1}+\mathrm{d}\sigma_0+\mathrm{d}\sigma_{-1}}{3},\notag\\
\mathrm{d}\sigma_{U,LL}&=\frac{\mathrm{d}\sigma_{+1}+\mathrm{d}\sigma_{-1}-2\mathrm{d}\sigma_0}{3},
\label{eq:ll-combination}
\end{align}
with beam helicities averaged. The first combination removes both vector and tensor polarization; the second cancels the unpolarized contribution and vector polarization. An $LT$ response can be isolated by reversing $\bm S_{LT}$ at fixed $S_{LL}$ and $S_{TT}$, whereas a rotation $\phi_{TT}\to\phi_{TT}+\pi/2$ reverses $S_{TT}^{ij}$. In practice, these combinations may be formed from several physical target states. The beam-helicity difference isolates the $L$ responses, with the actual beam-polarization magnitude divided out.

We normalize each tensor-polarized contribution to the azimuthally averaged unpolarized cross section after removing the corresponding polarization factors. For $i=LL,LT,TT$, let $\mathrm{d}\Phi_i$ denote the relevant angular integration measure, with
\begin{align}
\mathrm{d}\Phi_{LL}&=\mathrm{d}\phi_T,\notag\\
\mathrm{d}\Phi_{LT}&=\mathrm{d}\phi_T\,\mathrm{d}\phi_{LT},\notag\\
\mathrm{d}\Phi_{TT}&=\mathrm{d}\phi_T\,\mathrm{d}\phi_{TT}\,.\label{eq:angular-measures}
\end{align}
The angular ranges are $0\leq\phi_T,\phi_{LT}<2\pi$ and $0\leq\phi_{TT}<\pi$. The normalized azimuthal moments are defined as
\begin{align}
\widetilde A_{a,i}^{W}&=\frac{\displaystyle\int\mathrm{d}\Phi_i}{\displaystyle\int\mathrm{d}\Phi_i\,W^2}\frac{\displaystyle\int\mathrm{d}\Phi_i\,W\,\mathrm{d}\sigma_{a,i}}{\displaystyle\int\mathrm{d}\Phi_i\,\mathrm{d}\sigma_{UU}}\,,\quad a=U,L\,.\label{eq:moments}
\end{align}
For $W=1$, the normalization factor is unity, whereas for the sine and cosine harmonics considered here it equals 2. The azimuthally averaged unpolarized cross section is $\mathcal N A_0 f_1^g$. Different $LT$ harmonics can be separated by resolving the target orientation; at fixed $\phi_{LT}=0$, the two rank-one cosine weights coincide.

The two $LL$ observables are
\begin{align}
\widetilde A_{U,LL}&=\frac{f_{1LL}^g}{f_1^g},\label{eq:all0}\\
\widetilde A_{U,LL}^{\cos2\phi_T}&=\frac{q_T^2}{M_D^2}\frac{B_0}{A_0}
\frac{h_{1LL}^{\perp g}}{f_1^g}.
\label{eq:all2}
\end{align}
The four $LT$ observables are
\begin{align}
\widetilde A_{U,LT}^{\cos(\phi_T-\phi_{LT})}
&=\frac{q_T}{M_D}\frac{f_{1LT}^g+h_{1LT}^g}{f_1^g},\label{eq:alt-minus}\\
\widetilde A_{L,LT}^{\sin(\phi_T-\phi_{LT})}
&=\frac{q_T}{M_D}\frac{C_0}{A_0}\frac{g_{1LT}^g}{f_1^g},\label{eq:alt-sin}\\
\widetilde A_{U,LT}^{\cos(\phi_T+\phi_{LT})}
&=\frac{2q_T}{M_D}\frac{B_0}{A_0}\frac{h_{1LT}^g}{f_1^g},\label{eq:alt-plus}\\
\widetilde A_{U,LT}^{\cos(3\phi_T-\phi_{LT})}
&=-\frac{q_T^3}{2M_D^3}\frac{B_0}{A_0}\frac{h_{1LT}^{\perp g}}{f_1^g}.
\label{eq:alt-three}
\end{align}
The four $TT$ observables are
\begin{align}
\widetilde A_{U,TT}^{\cos2(\phi_T-\phi_{TT})}
&=\frac{q_T^2}{M_D^2}\frac{f_{1TT}^g-h_{1TT}^{\perp g}}{f_1^g},\label{eq:att-minus}\\
\widetilde A_{L,TT}^{\sin2(\phi_T-\phi_{TT})}
&=\frac{q_T^2}{M_D^2}\frac{C_0}{A_0}\frac{g_{1TT}^g}{f_1^g},\label{eq:att-sin}\\
\widetilde A_{U,TT}^{\cos2\phi_{TT}}
&=2\frac{B_0}{A_0}\frac{h_{1TT}^g}{f_1^g},\label{eq:att-zero}\\
\widetilde A_{U,TT}^{\cos(4\phi_T-2\phi_{TT})}
&=\frac{q_T^4}{2M_D^4}\frac{B_0}{A_0}\frac{h_{1TT}^{\perp\perp g}}{f_1^g}.
\label{eq:att-four}
\end{align}
Equations~\eqref{eq:all0}--\eqref{eq:att-four} give ten normalized observables. They do not all determine individual gluon TMDs: the $\cos2(\phi_T-\phi_{TT})$ moment involves $f_{1TT}^g-h_{1TT}^{\perp g}$, while the $\cos(\phi_T-\phi_{LT})$ moment involves $f_{1LT}^g+h_{1LT}^g$. The two rank-one $LT$ cosine modulations allow $f_{1LT}^g$ and $h_{1LT}^g$ to be separated if the target orientation is resolved and $B_0\neq0$.

The physical cross-section contributions are obtained by restoring the factors $S_{LL}$, $s_{LT}$, or $s_{TT}$ and, for the beam-helicity-dependent terms, $\lambda_e$. In particular, an individual normalized coefficient need not lie between $-1$ and $1$: positivity constraints apply to the complete cross section for a physical spin-1 density matrix~\cite{Cotogno:2017puy}.

\section{Spectator-model input and numerical setup}
\label{sec:setup}

We use the spin-1 spectator model for the T-even gluon distributions and its extension including the final-state interaction for $g_{1LT}^g$ and $g_{1TT}^g$~\cite{Xie:2026uja,Xie:2026zjn}. The model treats the residual system as an on-shell spin-1 spectator with a continuous mass $M_X$. Each distribution is obtained from
\begin{align}
F^g(x,k_T^2;Q_0)=\int_{M_D}^{\infty}\mathrm{d} M_X\,
\rho(M_X)F^g(x,k_T^2;M_X,Q_0).
\label{eq:spectral-integral}
\end{align}
The effective deuteron-gluon-spectator vertex contains three form factors. As described in Ref.~\cite{Xie:2026uja}, the model parameters were determined by fitting the transverse-momentum-integrated unpolarized gluon distribution to the nNNPDF1.0 input at $Q_0=2\,\mathrm{GeV}$. The gluon momentum fraction in the present calculation is defined relative to the entire deuteron, as in Eq.~\eqref{eq:xg}; the relation to the nucleon-level momentum fraction used in the input PDF fit is specified in Ref.~\cite{Xie:2026uja}. The analytic TMD expressions and the one-gluon-exchange contributions to the T-odd distributions are given in Refs.~\cite{Xie:2026uja,Xie:2026zjn}. We evaluate the spectator-model TMDs directly at $(x_g,q_T^2)$.

We use the 100 parameter replicas of the spectator model. For each replica, the numerator and denominator of an observable are evaluated using the same parameter set before forming their ratio. Writing $F_i$ for the relevant gluon TMD or TMD combination and $K_i$ for its kinematic prefactor in Eqs.~\eqref{eq:all0}--\eqref{eq:att-four}, we obtain
\begin{align}
\widetilde A_i^{(r)}=K_i\frac{F_i^{(r)}(x_g,q_T^2;Q_0)}
{f_1^{g,(r)}(x_g,q_T^2;Q_0)},\quad r=1,\ldots,100.
\label{eq:replica-ratio}
\end{align}
In the figures, the central curves correspond to replica 65. At each kinematic point, the uncertainty band extends from the 17th to the 84th value of the 100 replicas arranged in ascending order. The band reflects the variation of the fitted spectator-model parameters and is not an estimate of uncertainties from the LDMEs, TMD evolution, or higher-order corrections.

All TMDs are evaluated at the fixed input scale $Q_0=2\,\mathrm{GeV}$. Varying $Q^2$ changes the hard-scattering coefficients and the gluon momentum fraction $x_g$, but no TMD evolution is included. The overall QED and QCD couplings in the leading-order hard scattering cancel in the normalized observables. In the one-gluon-exchange calculation of the T-odd input, the strong coupling is fixed at $\alpha_s=0.3$~\cite{Xie:2026zjn}. The T-odd TMDs are evaluated with the normalization and $[+,+]$ gauge-link convention specified in Sec.~\ref{sec:formalism}.

The masses and reference collision energy are
\begin{align}
M_D&=1.876\,\mathrm{GeV},\qquad M_\psi=3.1\,\mathrm{GeV},\notag\\
m_c&=M_\psi/2,\qquad s_{eD}=4050\,\mathrm{GeV}^2.
\label{eq:setup-mass}
\end{align}
We adopt $\sqrt{s_{eN}}=45\,\mathrm{GeV}$ as a representative electron-nucleon center-of-mass energy at the EIC~\cite{Accardi:2012qut}. Using $s_{eD}\simeq 2s_{eN}$, we obtain $\sqrt{s_{eD}}\simeq63.64\,\mathrm{GeV}$, which is used for the $q_T$, $y$, and $Q^2$ scans. Throughout, $s_{eD}$ denotes the electron--deuteron invariant, consistent with the momentum fraction being defined relative to the whole deuteron in Eq.~\eqref{eq:xg}.

We adopt the positive leading-order SV matrix elements~\cite{Sharma:2012dy}, also used in Ref.~\cite{liu:2023unl},
\begin{align}
\mathcal O_S=\frac{\mathcal O_P}{m_c^2}=1.8\times10^{-2}\,\mathrm{GeV}^3.
\label{eq:sv}
\end{align}
Their common normalization cancels from $B_0/A_0$ and $C_0/A_0$, whereas the relative $S$- and $P$-wave contributions remain relevant. The calculation describes direct $J/\psi$ production in the specified color-octet channels. Feed-down from heavier charmonia, diffractive production, and other backgrounds are not included.

For the transverse-momentum scan we choose
\begin{align}
Q^2&=1,\ 5,\ 10\,\mathrm{GeV}^2,\qquad y=0.5,\notag\\
0.05&\leq q_T\leq1\,\mathrm{GeV}.
\label{eq:qt-scan}
\end{align}
The three virtualities correspond to $x_g\simeq0.00524$, $0.00721$, and $0.00968$, respectively. For the $y$ scan, we use the same values of $Q^2$ with $q_T=0.3\,\mathrm{GeV}$ and $0.1\leq y\leq1$. For the $Q^2$ scan, we take $q_T=0.3\,\mathrm{GeV}$, $y=0.3$, $0.5$, and $0.7$, and $4\leq Q^2\leq25\,\mathrm{GeV}^2$. The three settings are shown by blue solid, red dashed, and green dotted curves, respectively, with bands of the same colors. The limit $y\to1$ is shown for comparison. At $Q^2=1\,\mathrm{GeV}^2$, the region $q_T\sim1\,\mathrm{GeV}$ lies outside the strict condition $q_T\ll Q$; the corresponding results should be viewed as an extrapolation of the leading-power approximation.

The collision-energy scan is performed at
\begin{align}
Q^2=10\,\mathrm{GeV}^2,\qquad y=0.5,\qquad q_T=0.3\,\mathrm{GeV}.
\label{eq:s-scan}
\end{align}
Following Ref.~\cite{Mukherjee:2016qxa}, we consider four reference electron-nucleon center-of-mass energies, $\sqrt{s_{eN}}=4.7$, $7.2$, $17.33$, and $45\,\mathrm{GeV}$, motivated by the kinematics of JLab, HERMES, COMPASS, and the EIC, respectively. The COMPASS value corresponds to a muon-nucleon collision energy and is used here only as a numerical benchmark. Using $s_{eD}\simeq2s_{eN}$, the corresponding electron-deuteron energies are listed in Table~\ref{tab:energies}. We also calculate nine intermediate points, giving thirteen energies over $6.65\lesssim\sqrt{s_{eD}}\lesssim63.64\,\mathrm{GeV}$. The highlighted points in the energy plots are model calculations at these reference energies, not experimental measurements; no facility-specific acceptance is imposed.

\begin{table}[tb]
\centering
\caption{Reference electron-nucleon and electron-deuteron energies and the corresponding gluon momentum fractions, using $s_{eD}=2s_{eN}$. The first two columns are in GeV. The lowest-energy entry is below the hadronic production threshold at the kinematics of Eq.~\eqref{eq:s-scan}; the second entry probes the large-$x_g$ region where the spectator-model estimates are less constrained.}
\label{tab:energies}
\begin{tabular}{ccc}
\toprule
$\sqrt{s_{eN}}$ & $\sqrt{s_{eD}}$ & $x_g$ \\
\midrule
4.70 & 6.65 & 0.8877 \\
7.20 & 10.18 & 0.3783 \\
17.33 & 24.51 & 0.06530 \\
45.00 & 63.64 & 0.009684 \\
\bottomrule
\end{tabular}
\end{table}

The virtualities in Eq.~\eqref{eq:qt-scan} sample markedly different hard-scattering weights. The SV relation $\mathcal O_P=M_\psi^2\mathcal O_S/4$ reduces Eqs.~\eqref{eq:a1}--\eqref{eq:c} to
\begin{align}
\mathcal A_1&=4\mathcal O_S\frac{2M_\psi^2+Q^2}{M_\psi^2+Q^2},\notag\\
\mathcal A_2&=8\mathcal O_S\frac{M_\psi^2Q^2}{(M_\psi^2+Q^2)^2},\notag\\
\mathcal B&=2\mathcal O_S\frac{M_\psi^2-Q^2}{M_\psi^2+Q^2},\notag\\
\mathcal C&=4\mathcal O_S\frac{Q^2}{M_\psi^2+Q^2}.
\label{eq:sv-hard}
\end{align}
Thus $\mathcal B$ changes sign at $Q^2=M_\psi^2=9.61\,\mathrm{GeV}^2$. The setting $Q^2=10\,\mathrm{GeV}^2$ and $y=0.5$, used in the energy scan, gives
\begin{align}
\frac{B_0}{A_0}\simeq-2.861\times10^{-3},\qquad
\frac{C_0}{A_0}\simeq-0.2201.
\label{eq:hard-benchmark}
\end{align}
Consequently, the moments proportional to $B_0$ are suppressed at $Q^2=10\,\mathrm{GeV}^2$, which is close to the zero of $\mathcal B$. In contrast, $C_0$ does not vanish there. The observables in Eqs.~\eqref{eq:all0}, \eqref{eq:alt-minus}, and \eqref{eq:att-minus} are independent of these ratios because their common hard-scattering coefficient cancels.

For a more general LDME ratio, the condition $\mathcal B=0$ becomes
\begin{align}
\frac{Q_{\rm zero}^2}{M_\psi^2}
=\frac{3r-1}{1+r},\qquad
r=\frac{4\mathcal O_P}{M_\psi^2\mathcal O_S},
\label{eq:zero-general}
\end{align}
when the right-hand side is positive and $1+r\neq0$. The position of this zero depends on the relative color-octet LDMEs rather than on the gluon TMDs.

\section{Numerical results and discussion}
\label{sec:results}

\begin{figure*}[t]
\centering
\includegraphics[width=0.485\textwidth]{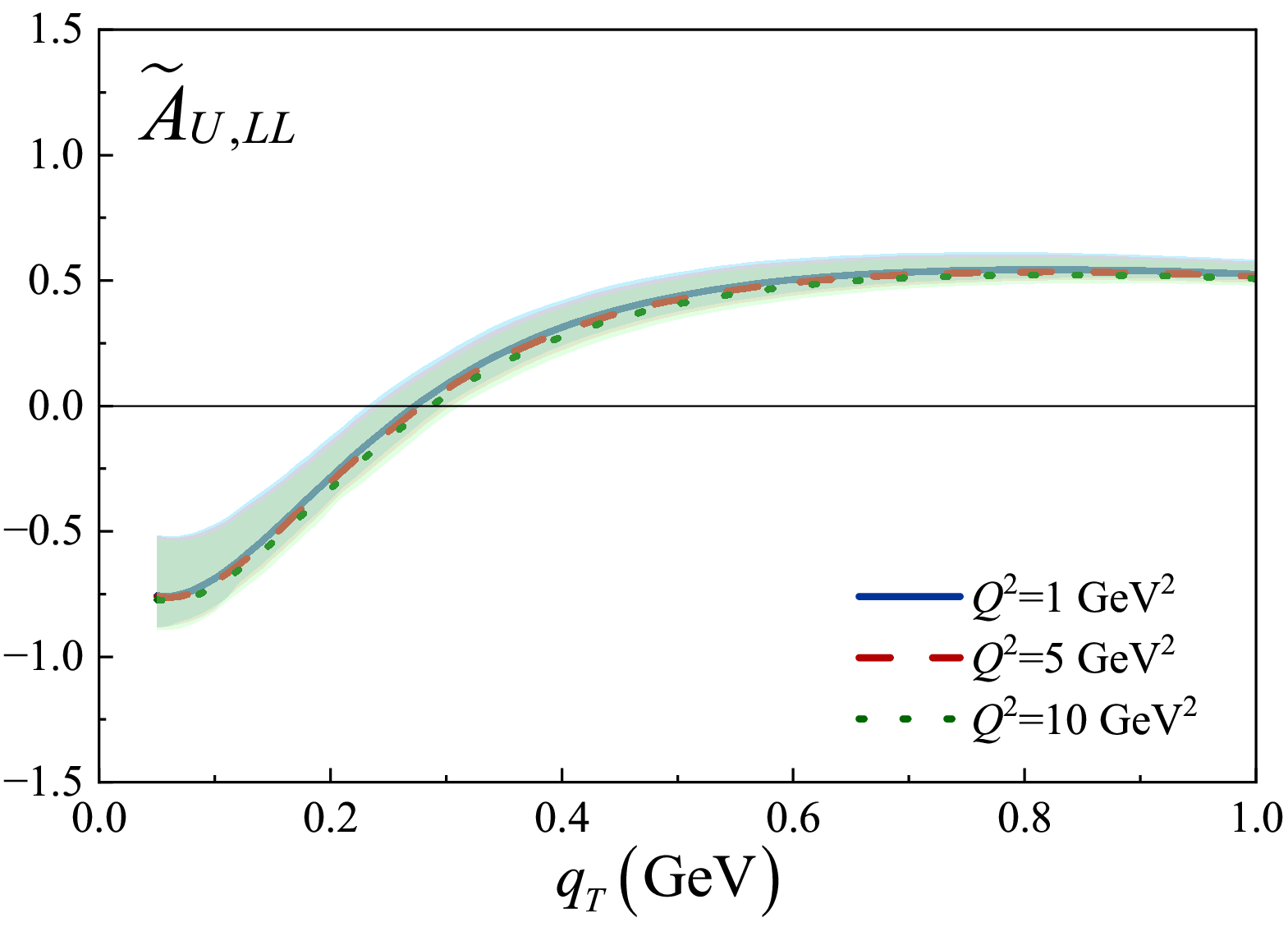}\hfill
\includegraphics[width=0.485\textwidth]{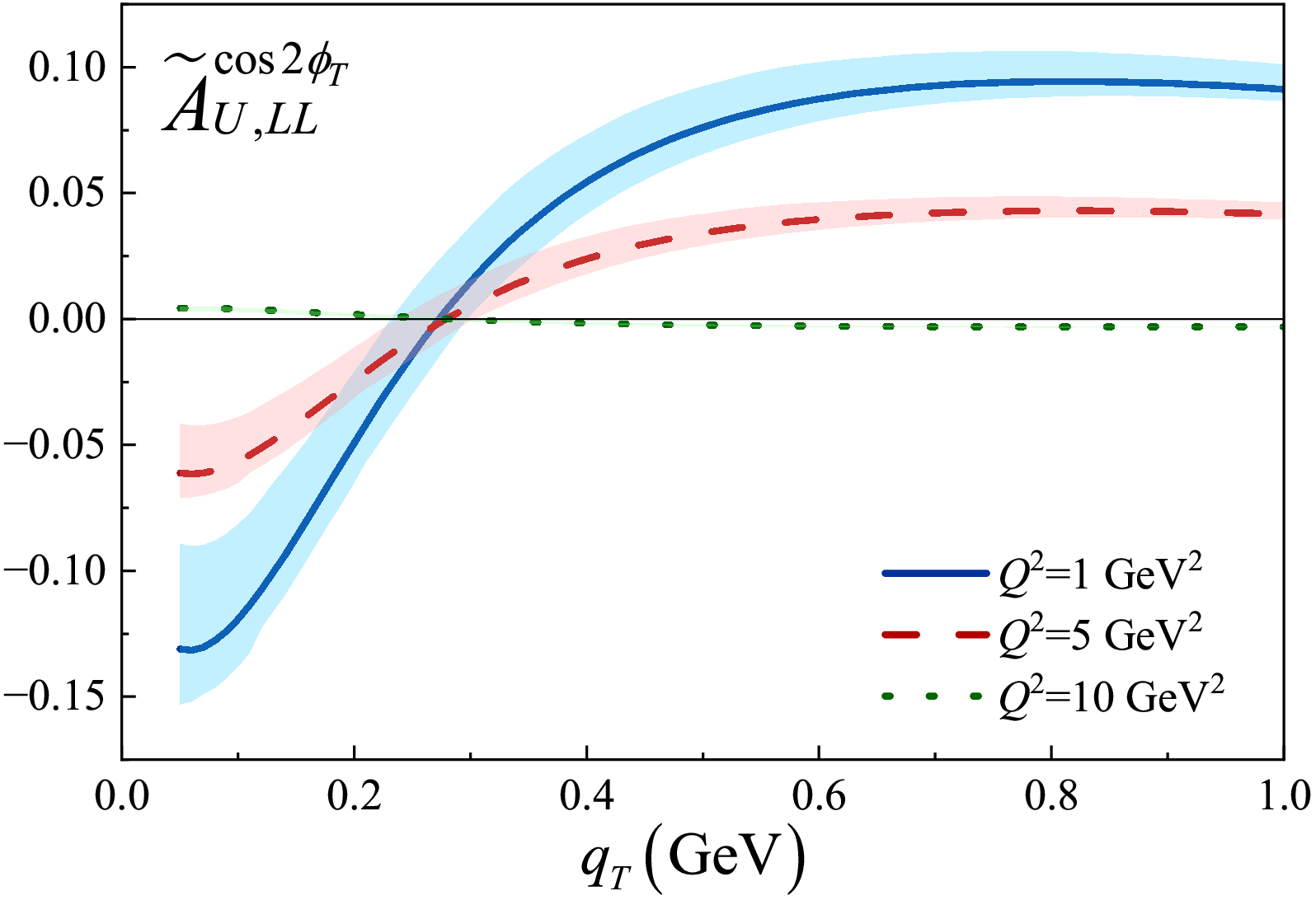}
\caption{Normalized $LL$ observables versus $q_T$ at $y=0.5$ and $s_{eD}=4050\,\mathrm{GeV}^2$. The left and right panels correspond to Eqs.~\eqref{eq:all0} and \eqref{eq:all2}. Blue solid, red dashed, and green dotted curves denote $Q^2=1$, $5$, and $10\,\mathrm{GeV}^2$, respectively. Central curves use replica 65, and the bands extend from the 17th to the 84th ordered values of the 100 replicas. Note the different vertical scales.}
\label{fig:ll}
\end{figure*}

\begin{figure*}[t]
\centering
\includegraphics[width=0.485\textwidth]{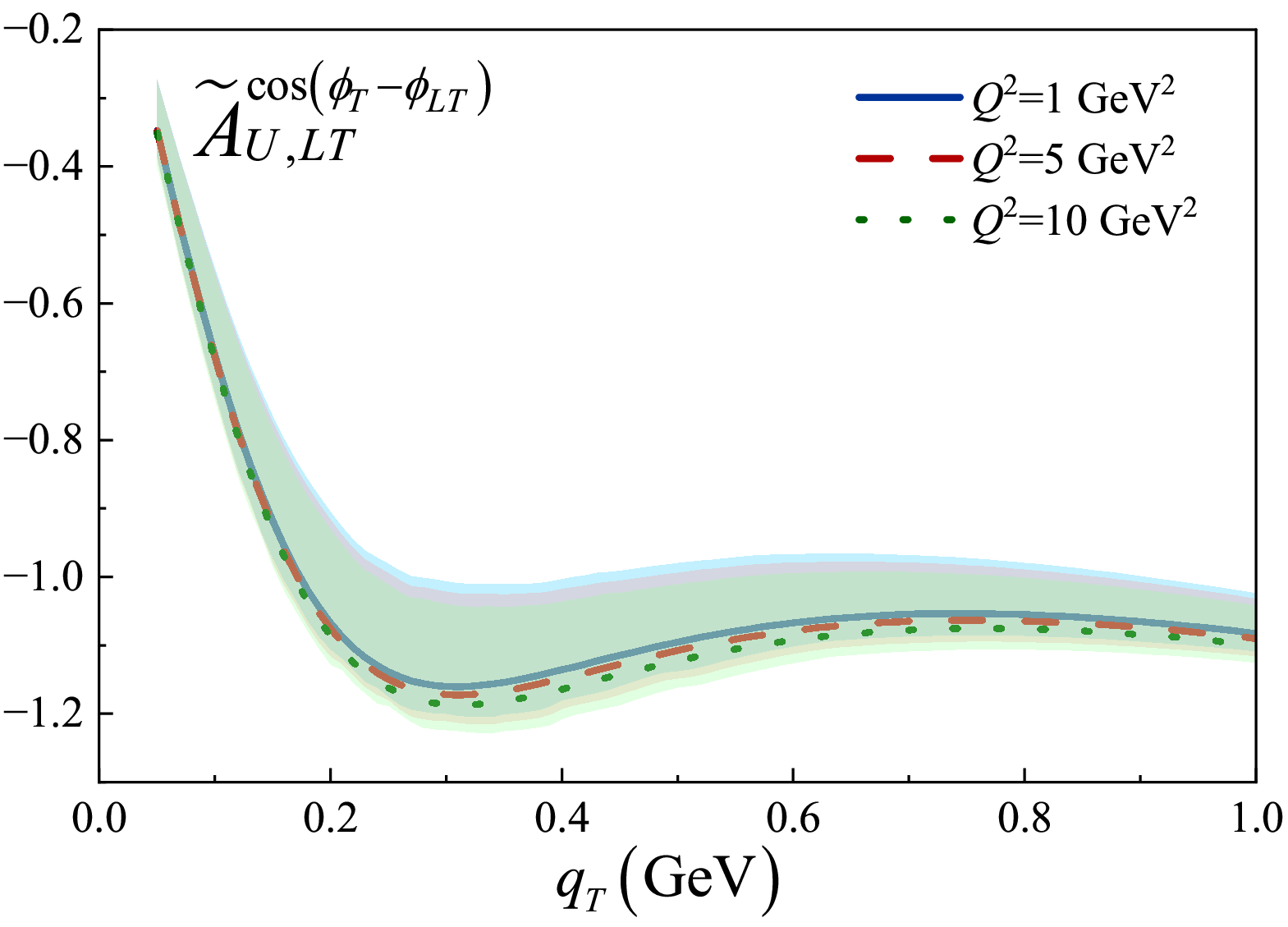}\hfill
\includegraphics[width=0.485\textwidth]{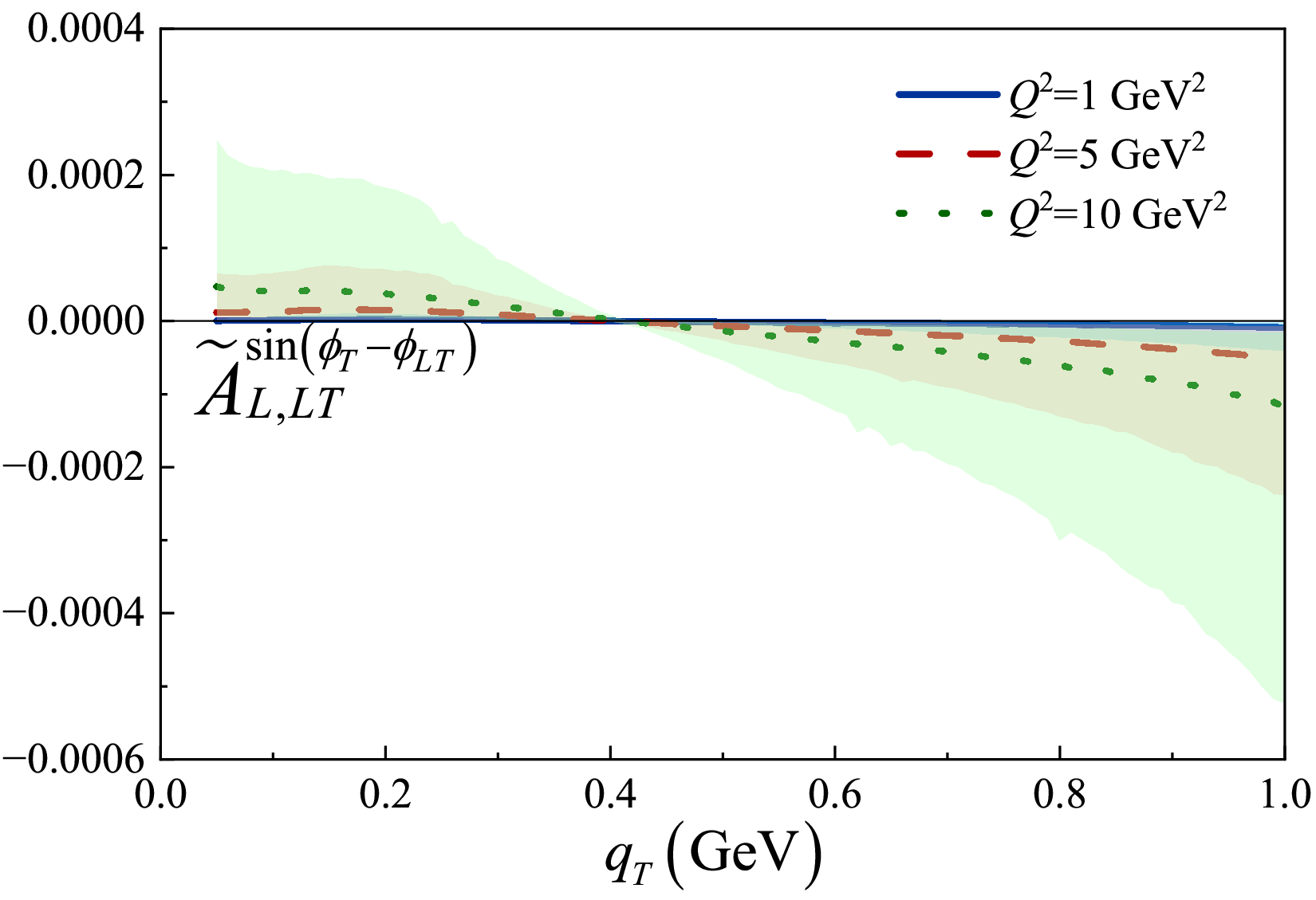}\\[5pt]
\includegraphics[width=0.485\textwidth]{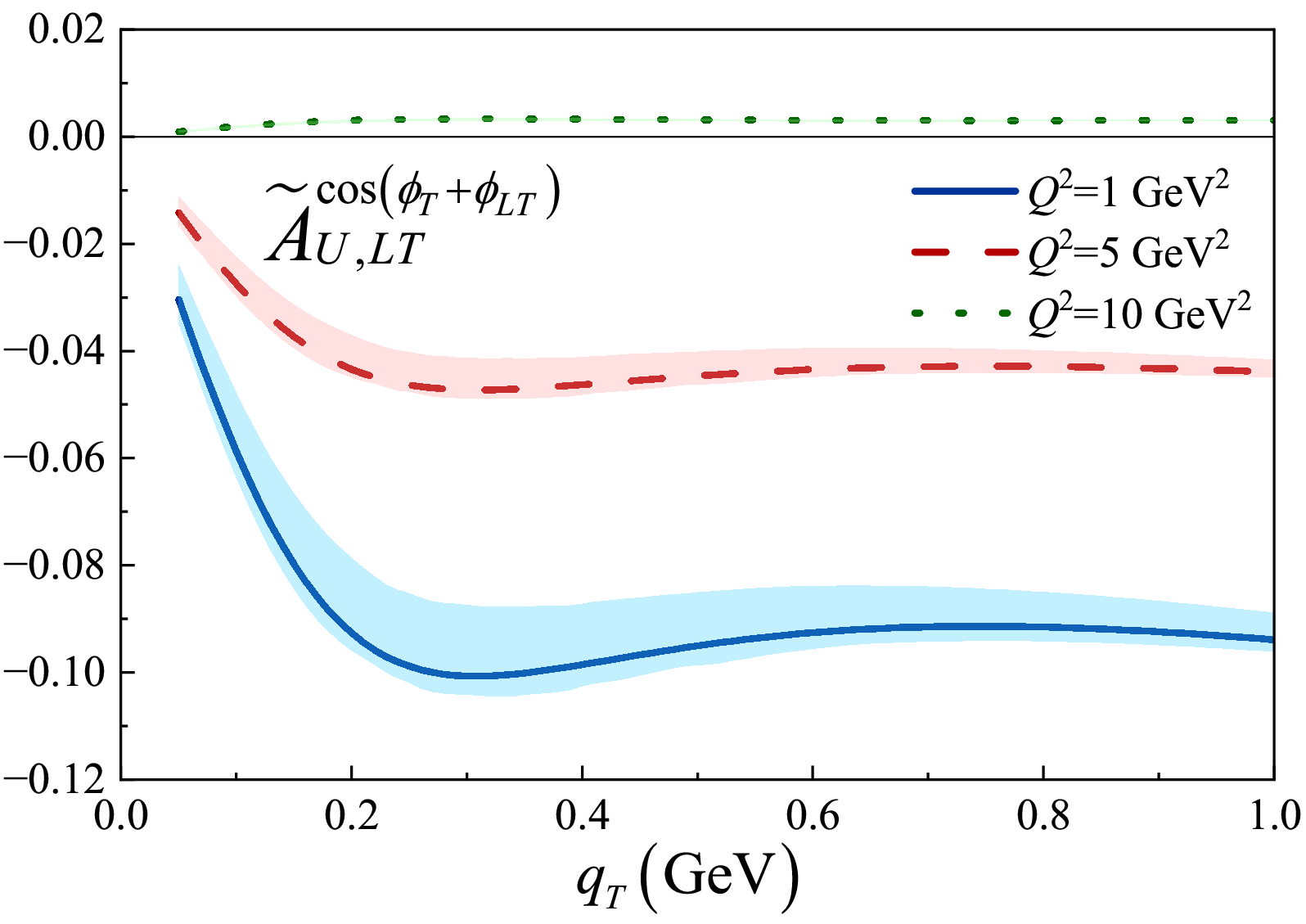}\hfill
\includegraphics[width=0.485\textwidth]{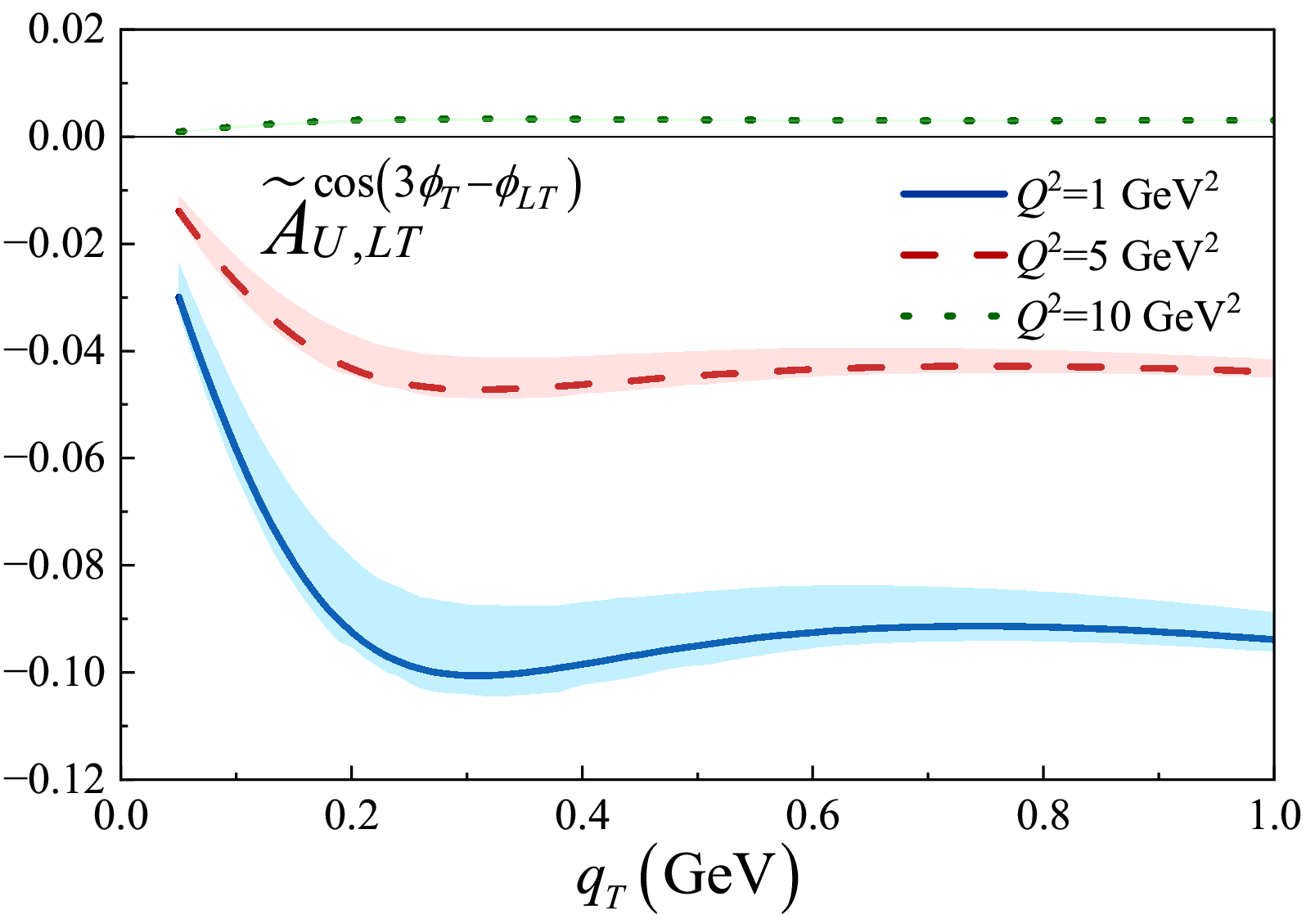}
\caption{Normalized $LT$ observables versus $q_T$, with the kinematics, line styles, and uncertainty bands of Fig.~\ref{fig:ll}. From left to right and top to bottom, the panels correspond to Eqs.~\eqref{eq:alt-minus}--\eqref{eq:alt-three} and involve $f_{1LT}^g+h_{1LT}^g$, $g_{1LT}^g$, $h_{1LT}^g$, and $h_{1LT}^{\perp g}$, respectively. The beam polarization is divided out of the sine moment.}
\label{fig:lt}
\end{figure*}

\begin{figure*}[t]
\centering
\includegraphics[width=0.485\textwidth]{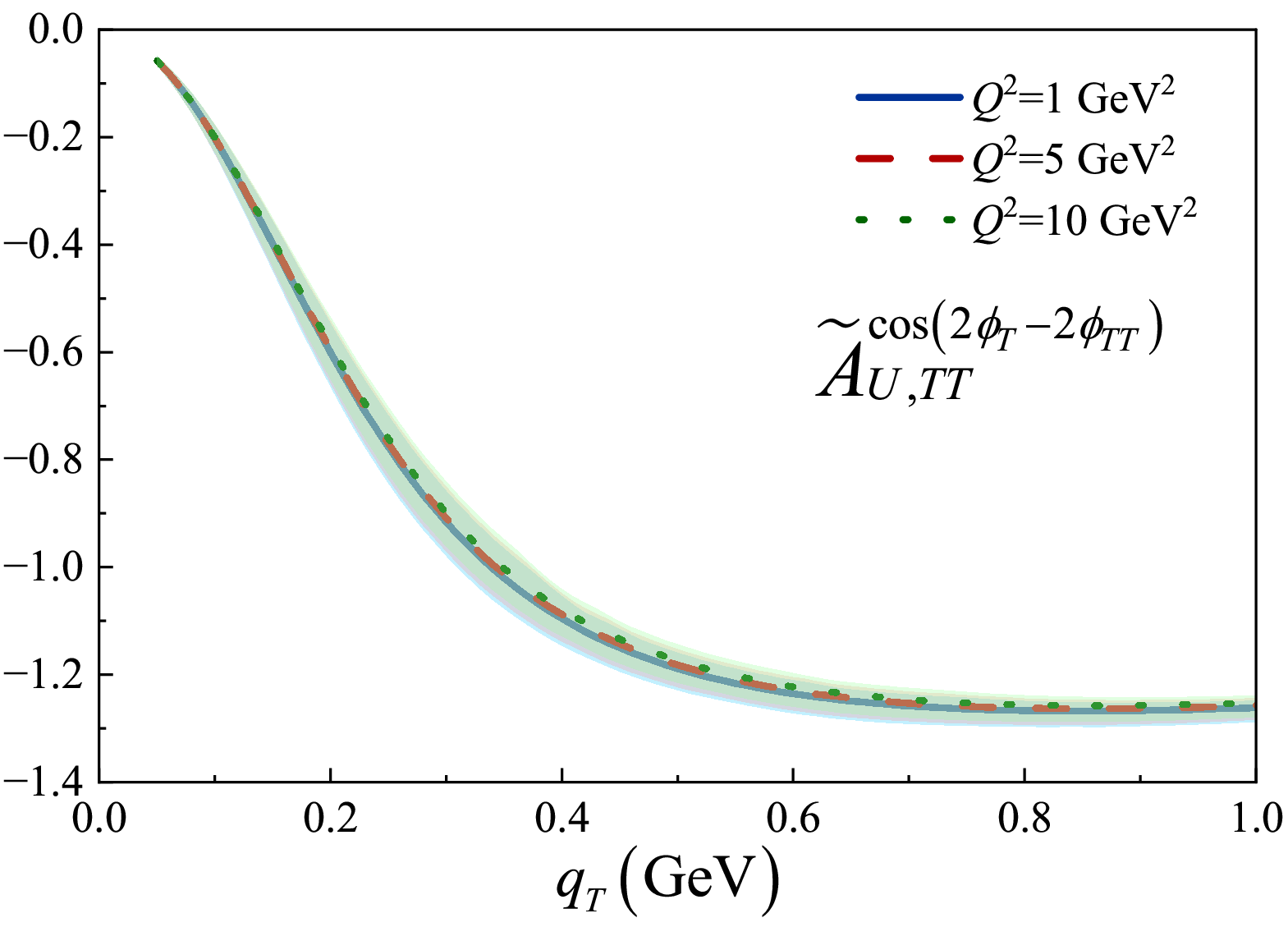}\hfill
\includegraphics[width=0.485\textwidth]{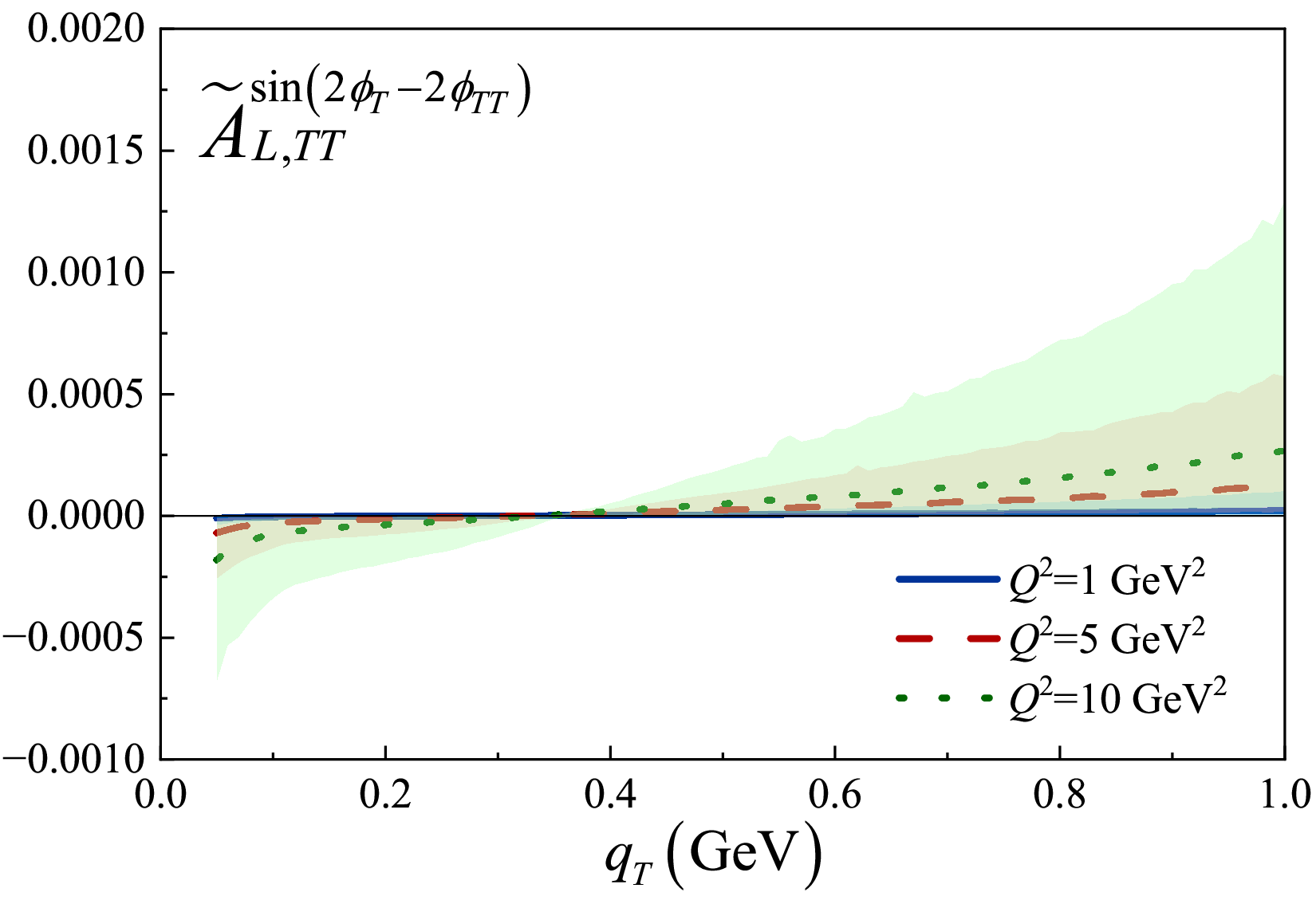}\\[5pt]
\includegraphics[width=0.485\textwidth]{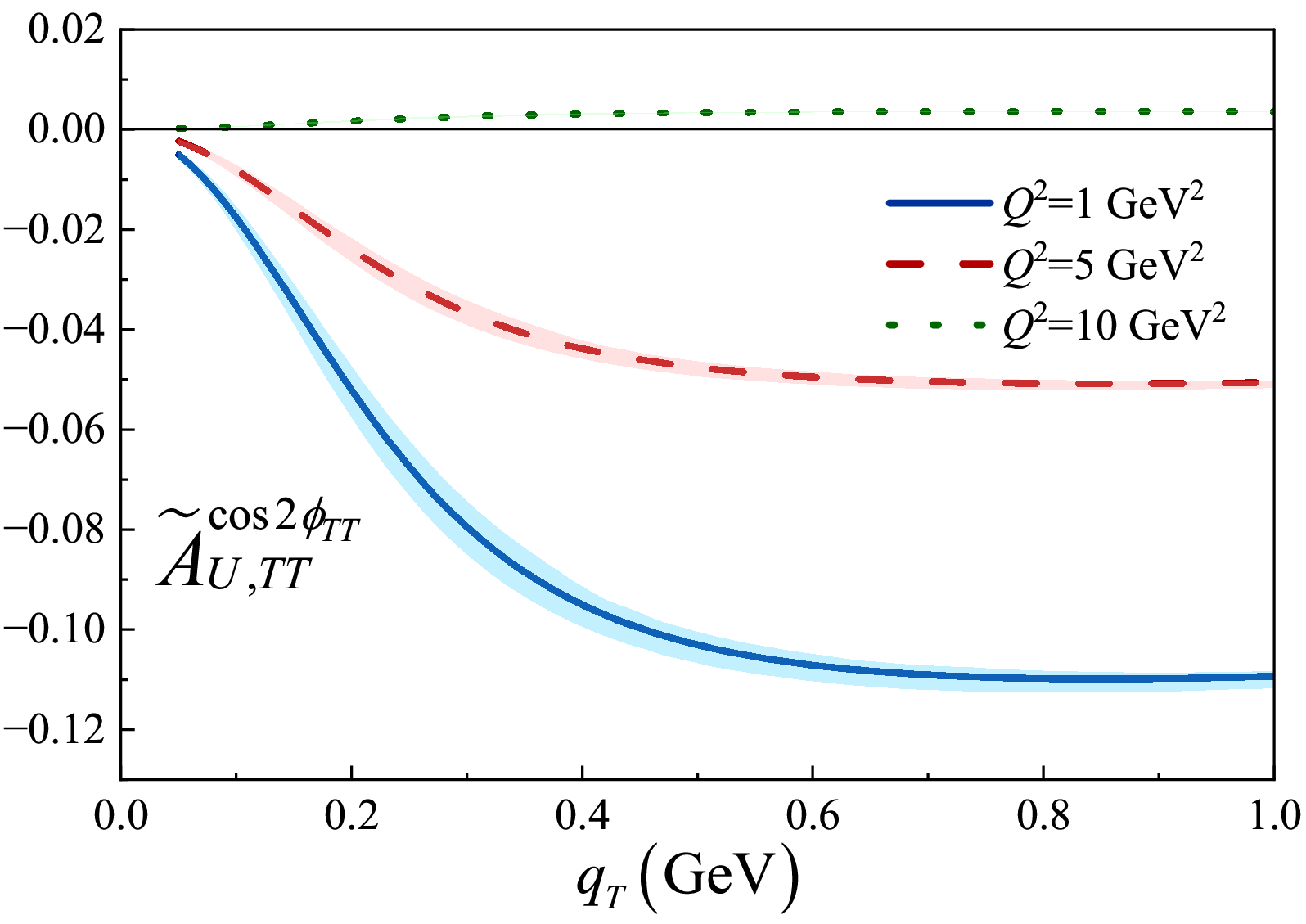}\hfill
\includegraphics[width=0.485\textwidth]{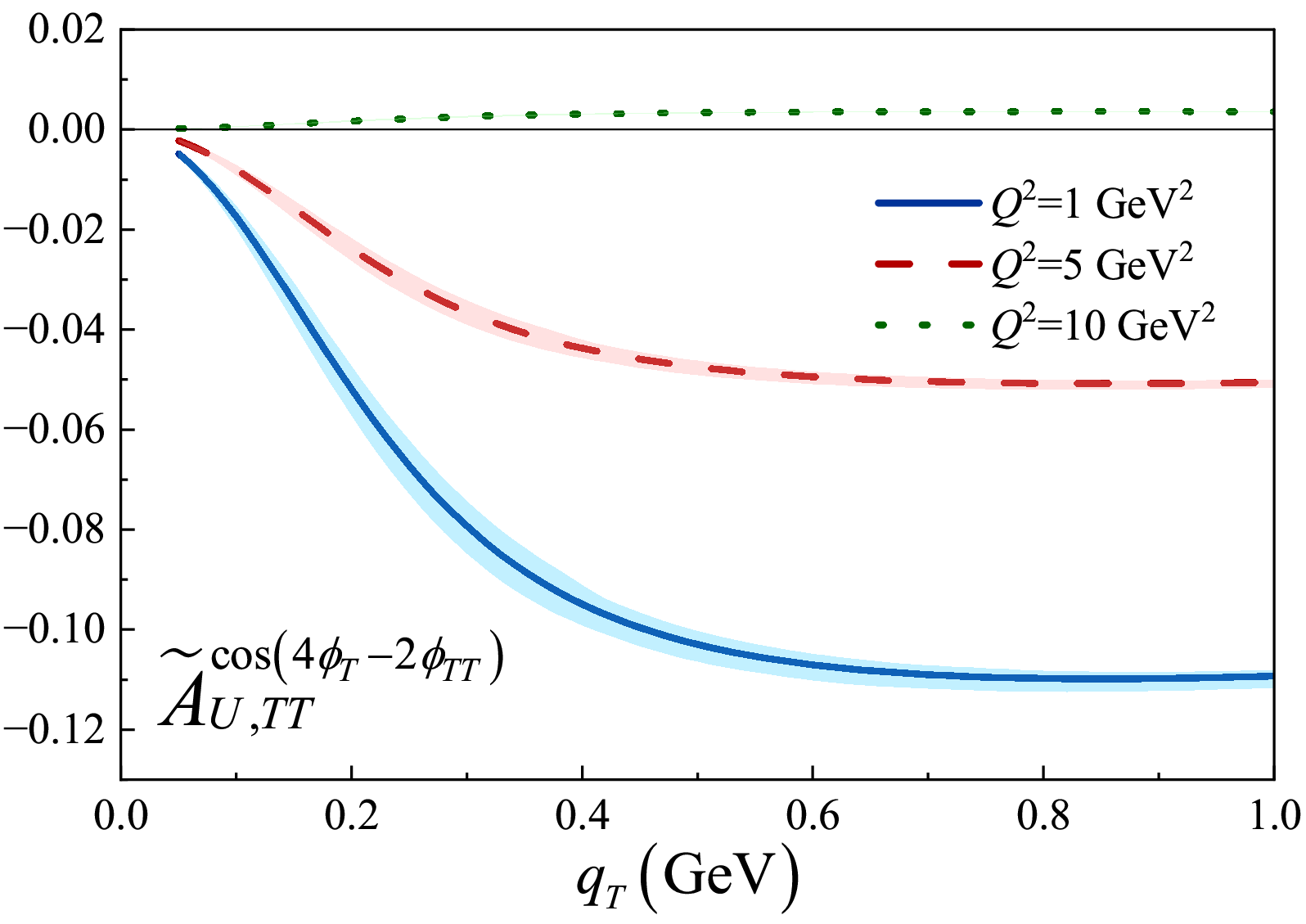}
\caption{Normalized $TT$ observables versus $q_T$, with the settings of Fig.~\ref{fig:ll}. From left to right and top to bottom, the panels correspond to Eqs.~\eqref{eq:att-minus}--\eqref{eq:att-four} and involve $f_{1TT}^g-h_{1TT}^{\perp g}$, $g_{1TT}^g$, $h_{1TT}^g$, and $h_{1TT}^{\perp\perp g}$, respectively. The sine moment requires a longitudinally polarized electron beam, whose polarization factor has been divided out.}
\label{fig:tt}
\end{figure*}

Figures~\ref{fig:ll}--\ref{fig:tt} show the $q_T$ dependence of all ten normalized observables at $Q^2=1$, $5$, and $10\,\mathrm{GeV}^2$. At fixed model scale, their dependence on $Q^2$ arises from the change in $x_g$ and, for some modulations, the ratios $B_0/A_0$ or $C_0/A_0$.

The $LL$ ratio $\widetilde A_{U,LL}=f_{1LL}^g/f_1^g$ changes sign near $q_T=0.3\,\mathrm{GeV}$, as shown in Fig.~\ref{fig:ll}. Its weak dependence on $Q^2$ follows from the similar values of the model ratio over $0.0052\lesssim x_g\lesssim0.0097$. The zero reflects a sign change in $f_{1LL}^g$ at fixed gluon momentum fraction.

The $\cos2\phi_T$ moment has a stronger dependence on $Q^2$ because it contains $B_0/A_0$. It is negative at small $q_T$ and positive at larger $q_T$ for $Q^2=1$ and $5\,\mathrm{GeV}^2$, while the signs are reversed and the magnitude is strongly suppressed at $Q^2=10\,\mathrm{GeV}^2$. At fixed $Q^2$, the zero as a function of $q_T$ comes from $h_{1LL}^{\perp g}$, whereas the overall sign reversal between virtualities is caused by the hard-scattering coefficient $\mathcal B$.

The $\cos(\phi_T-\phi_{LT})$ moment is negative throughout the plotted range and reaches approximately $-1.2$ near $q_T=0.3\,\mathrm{GeV}$. Its weak $Q^2$ dependence is consistent with Eq.~\eqref{eq:alt-minus}, in which the common hard-scattering coefficient cancels. As explained in Sec.~\ref{sec:formalism}, these results are normalized coefficients with the target-polarization factor removed.

The $\cos(\phi_T+\phi_{LT})$ and $\cos(3\phi_T-\phi_{LT})$ moments have nearly the same $q_T$ dependence, despite the different powers of $q_T$ in Eqs.~\eqref{eq:alt-plus} and \eqref{eq:alt-three}. Both are negative for $Q^2=1$ and $5\,\mathrm{GeV}^2$ and small and positive for $Q^2=10\,\mathrm{GeV}^2$. This behavior follows from the sign change of $\mathcal B$ and the relations between the corresponding TMDs in the spectator model, discussed below.

The beam-helicity-dependent $\sin(\phi_T-\phi_{LT})$ moment probes the T-odd distribution $g_{1LT}^g$ and is substantially smaller in magnitude. For $Q^2=10\,\mathrm{GeV}^2$, the central curve crosses zero around $q_T=0.4\,\mathrm{GeV}$ and reaches approximately $-1.1\times10^{-4}$ at $q_T=1\,\mathrm{GeV}$. The uncertainty band is broad relative to this central value. In our spectator model, $g_{1LT}^g$ depends sensitively on the vertex form factor $g_2$ and vanishes in the limit $g_2\to0$~\cite{Xie:2026zjn}.

The $\cos2(\phi_T-\phi_{TT})$ moment is negative and depends weakly on $Q^2$, consistent with the cancellation of the hard-scattering coefficient in Eq.~\eqref{eq:att-minus}. The two other $TT$ cosine moments have similar $q_T$ dependences; their sign changes between the lower virtualities and $Q^2=10\,\mathrm{GeV}^2$ follow from $\mathcal B$. The $\sin2(\phi_T-\phi_{TT})$ moment, which involves the T-odd distribution $g_{1TT}^g$, changes sign near $q_T=0.35$--$0.4\,\mathrm{GeV}$ and reaches approximately $2.6\times10^{-4}$ at $q_T=1\,\mathrm{GeV}$ for $Q^2=10\,\mathrm{GeV}^2$. Its replica band becomes wider at larger $q_T$. Like $g_{1LT}^g$, the distribution $g_{1TT}^g$ vanishes as $g_2\to0$ in our spectator model~\cite{Xie:2026zjn}.

The close agreement between the two $LT$ moments proportional to $B_0$, and likewise between the corresponding $TT$ moments, can be understood from the spectator-model relations. When the vertex form factor $g_2$ is set to zero, the T-even TMDs obey~\cite{Xie:2026uja}
\begin{align}
h_{1LT}^g&=-\frac{q_T^2}{4M_D^2}h_{1LT}^{\perp g},\notag\\
h_{1TT}^g&=\frac{q_T^4}{4M_D^4}h_{1TT}^{\perp\perp g}.
\label{eq:model-relations}
\end{align}
Inserting Eq.~\eqref{eq:model-relations} into Eqs.~\eqref{eq:alt-plus}, \eqref{eq:alt-three}, \eqref{eq:att-zero}, and \eqref{eq:att-four} yields identical moments in this limit. The near equality in Figs.~\ref{fig:lt} and \ref{fig:tt} is therefore a feature of the model input, not a model-independent relation. Measuring both angular modulations would help test the assumed TMD relations.

\begin{figure*}[t]
\centering
\includegraphics[width=0.485\textwidth]{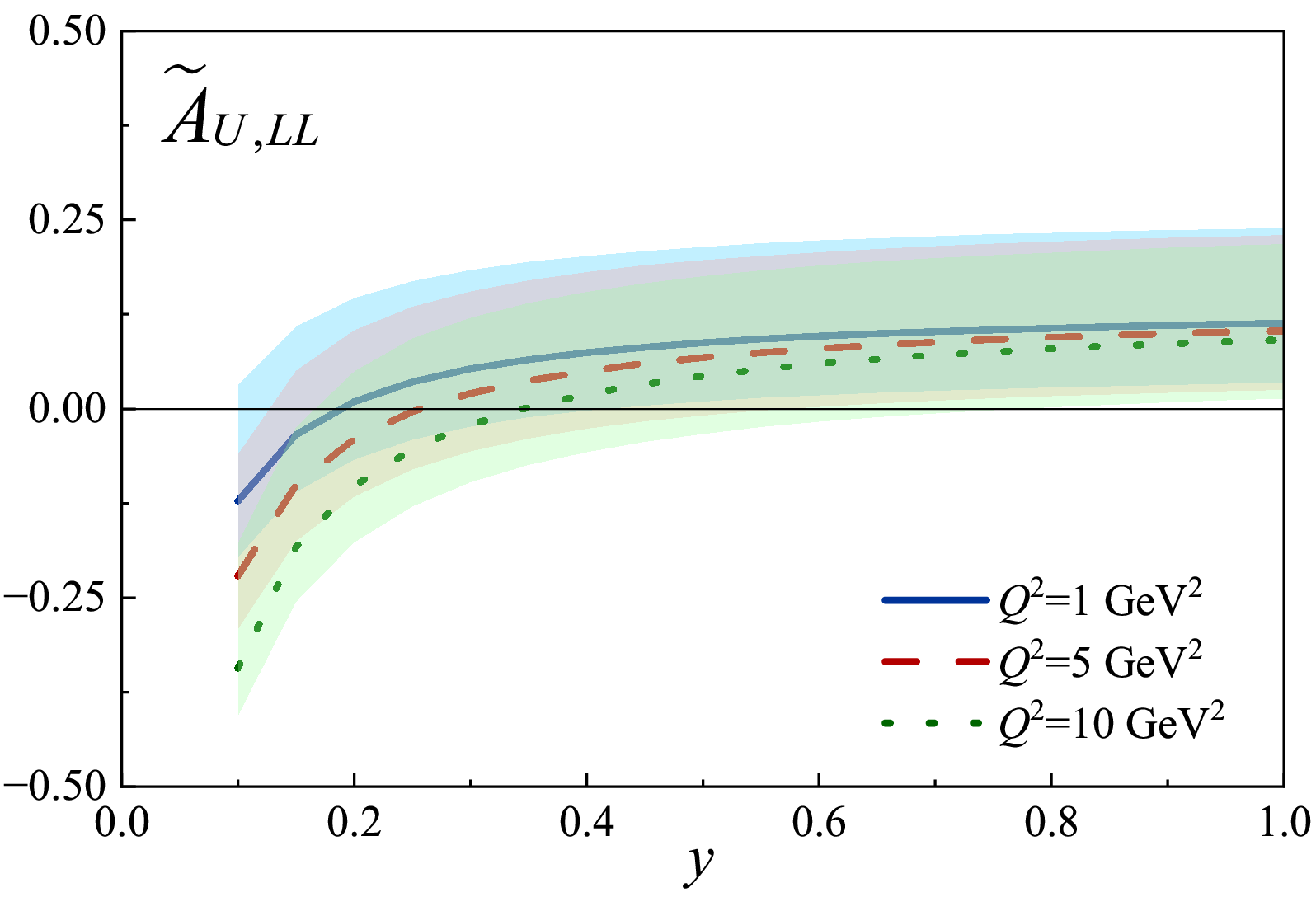}\hfill
\includegraphics[width=0.485\textwidth]{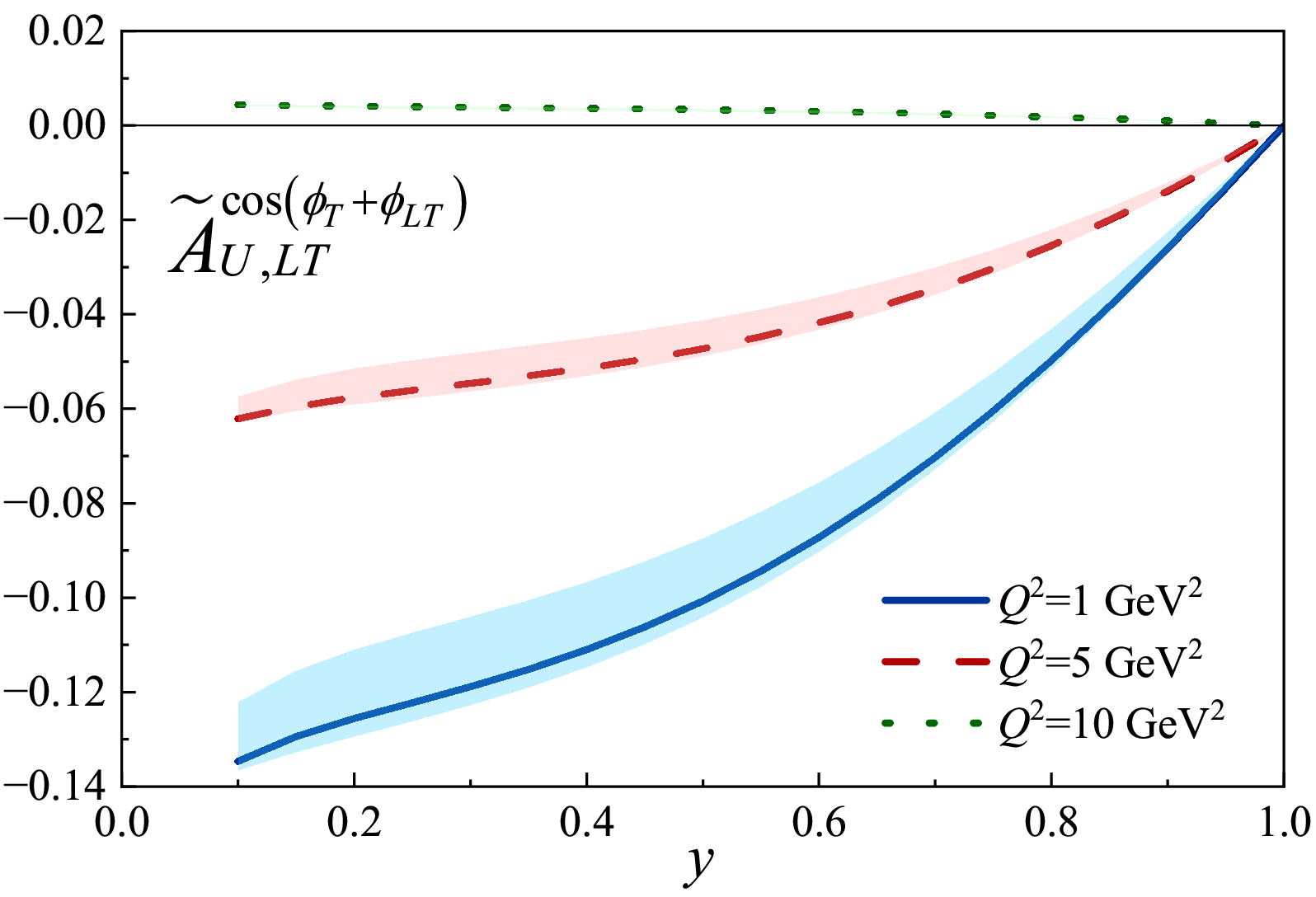}\\[5pt]
\includegraphics[width=0.485\textwidth]{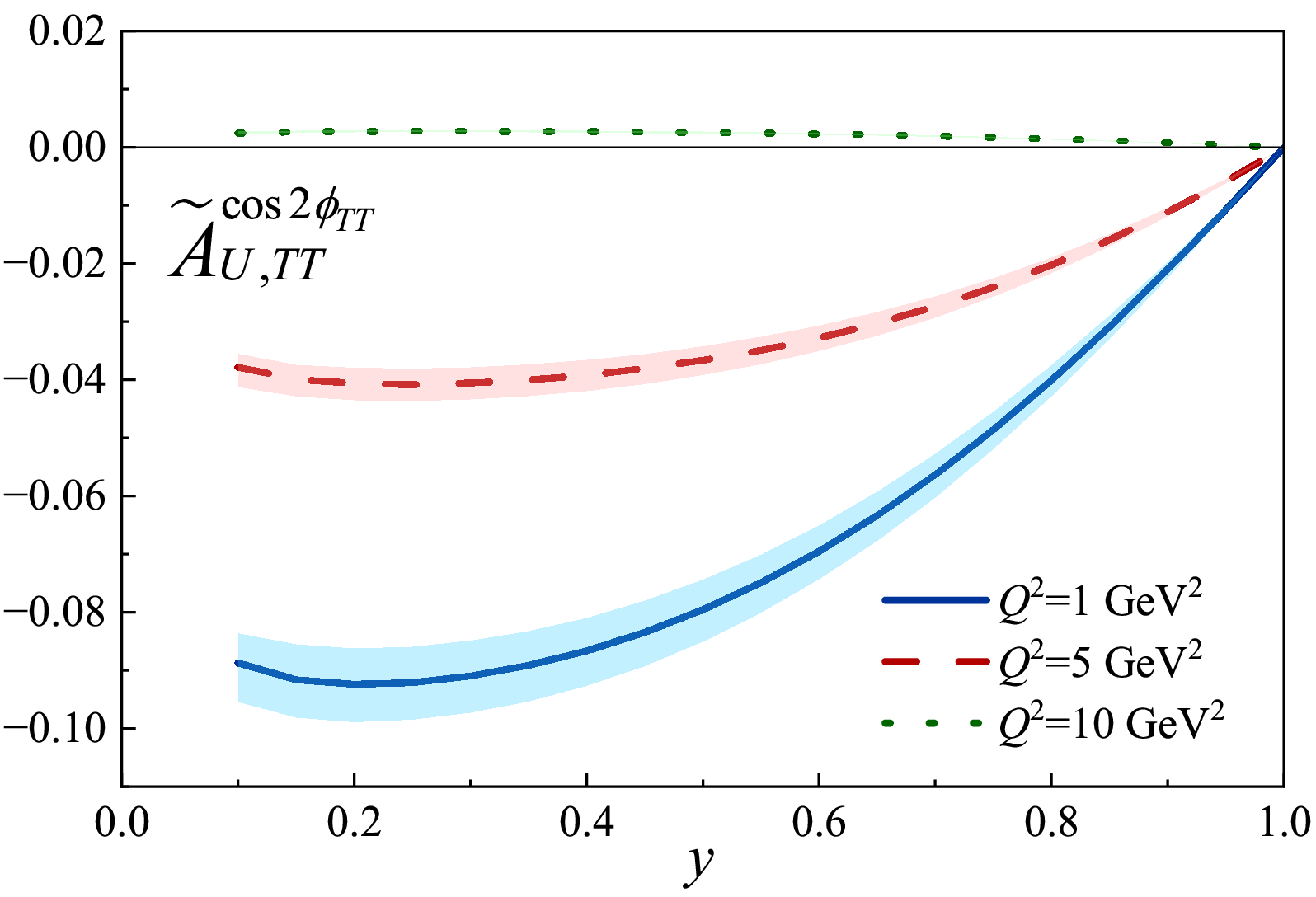}\hfill
\includegraphics[width=0.485\textwidth]{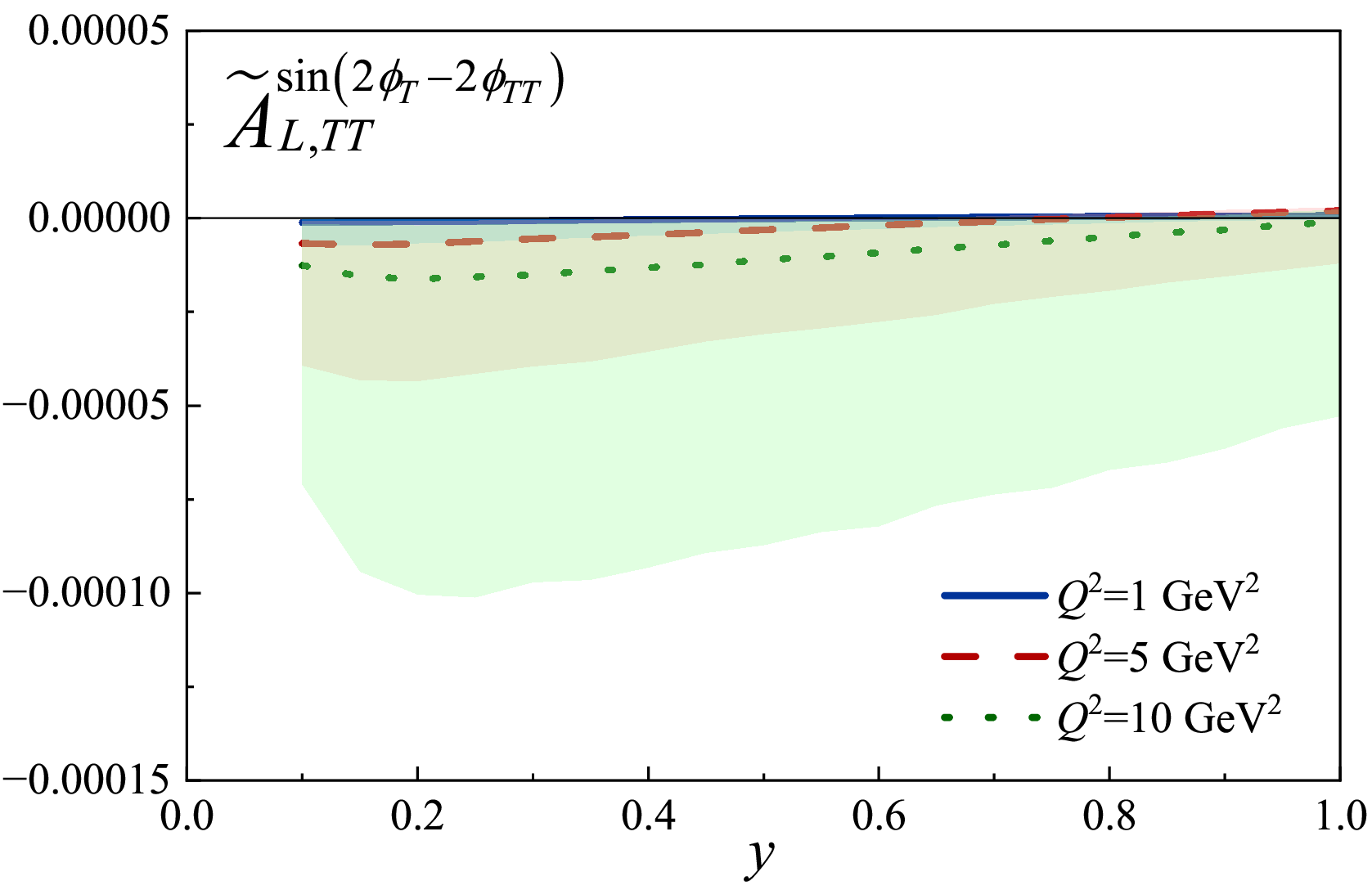}
\caption{Four representative normalized observables versus $y$ at $q_T=0.3\,\mathrm{GeV}$ and $s_{eD}=4050\,\mathrm{GeV}^2$. Blue solid, red dashed, and green dotted curves correspond to $Q^2=1$, $5$, and $10\,\mathrm{GeV}^2$, with the uncertainty bands described in Fig.~\ref{fig:ll}. From left to right and top to bottom, the panels involve $f_{1LL}^g$, $h_{1LT}^g$, $h_{1TT}^g$, and $g_{1TT}^g$, respectively. The endpoint $y=1$ is a formal limit.}
\label{fig:y}
\end{figure*}

We next consider the $y$, $Q^2$, and collision-energy dependences of four representative observables: $\widetilde A_{U,LL}$, $\widetilde A_{U,LT}^{\cos(\phi_T+\phi_{LT})}$, $\widetilde A_{U,TT}^{\cos2\phi_{TT}}$, and $\widetilde A_{L,TT}^{\sin2(\phi_T-\phi_{TT})}$. They are sensitive to $f_{1LL}^g$, $h_{1LT}^g$, $h_{1TT}^g$, and $g_{1TT}^g$, respectively, and differ in their dependence on the hard-scattering coefficients. Figure~\ref{fig:y} shows the $y$ dependence at three values of $Q^2$.

The $LL$ ratio in Fig.~\ref{fig:y} increases with $y$ and changes sign because the sampled $x_g=(M_\psi^2+Q^2)/(ys_{eD})$ decreases. If $f_{1LL}^g$ vanishes at $x_g=x_0$ for fixed $q_T$, the zero occurs at $y_0=(M_\psi^2+Q^2)/(s_{eD}x_0)$; this accounts for its shift toward larger $y$ as $Q^2$ increases. The overlapping uncertainty bands reflect the variation among model replicas.

Both linearly polarized gluon moments are negative for $Q^2=1$ and $5\,\mathrm{GeV}^2$ and positive but strongly suppressed for $Q^2=10\,\mathrm{GeV}^2$. They vanish as $y\to1$ because $B_0=(1-y)\mathcal B$, provided the TMD ratios remain finite. Away from this endpoint, their shapes reflect both the leptonic factor and the $x_g$ dependence of the gluon TMDs.

The $g_{1TT}^g$ sine moment is much smaller than the two cosine moments. Unlike $B_0$, the coefficient $C_0=-y(2-y)\mathcal C$ need not vanish at $y=1$. Its small magnitude in the plotted range therefore arises from the TMD ratio, together with the model parameters, rather than from an endpoint zero of the hard-scattering coefficient.

\begin{figure*}[t]
\centering
\includegraphics[width=0.485\textwidth]{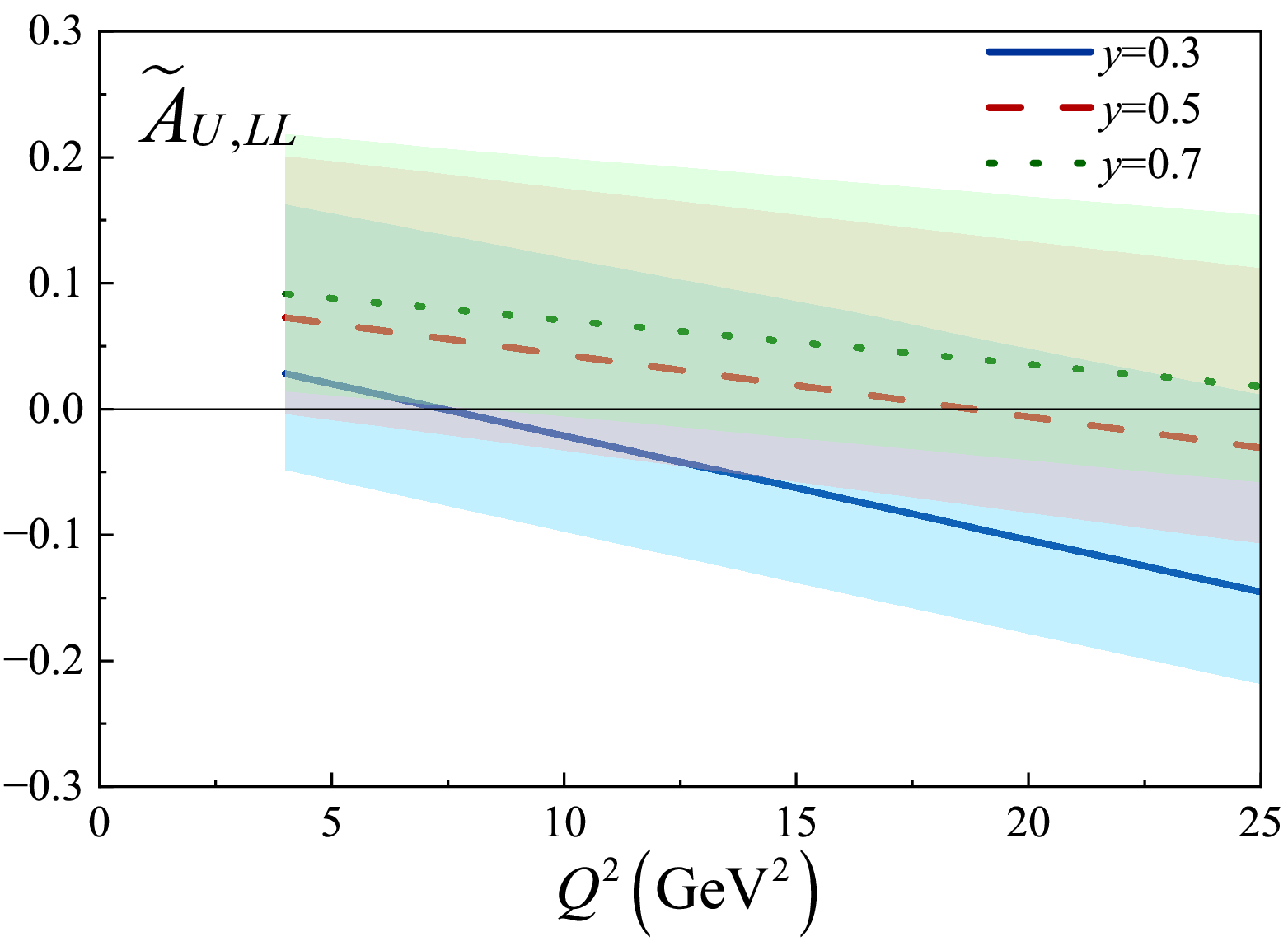}\hfill
\includegraphics[width=0.485\textwidth]{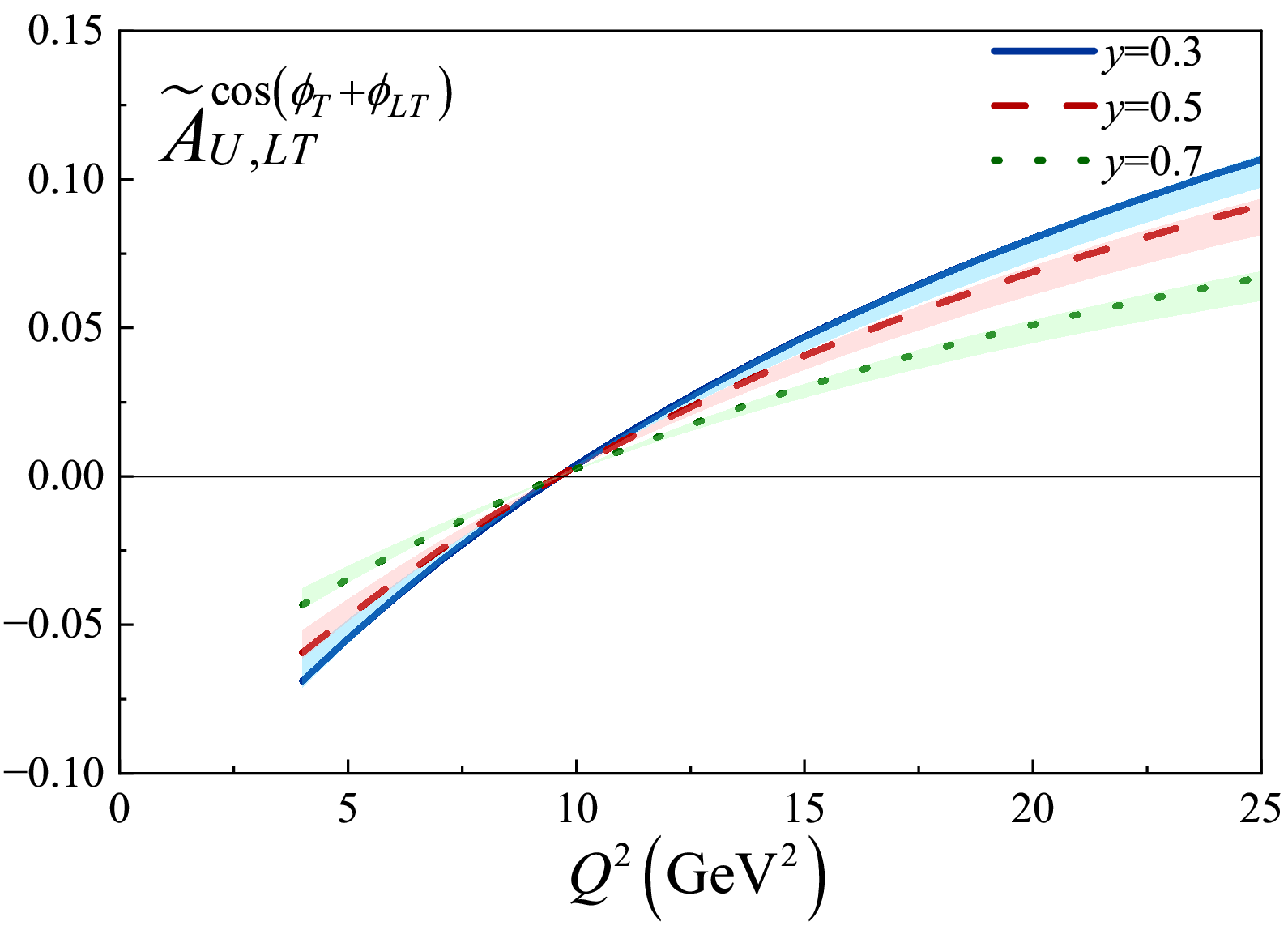}\\[5pt]
\includegraphics[width=0.485\textwidth]{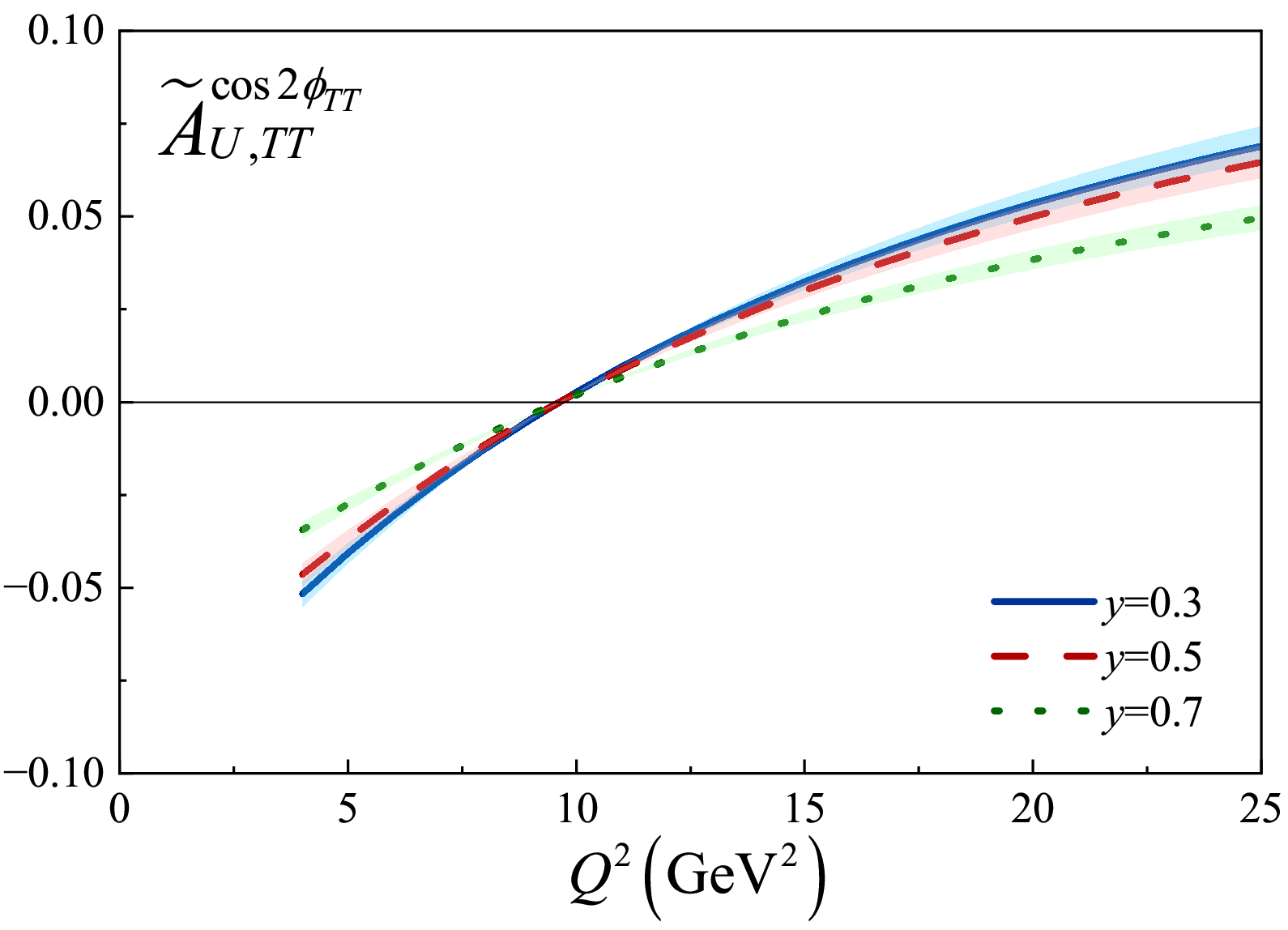}\hfill
\includegraphics[width=0.485\textwidth]{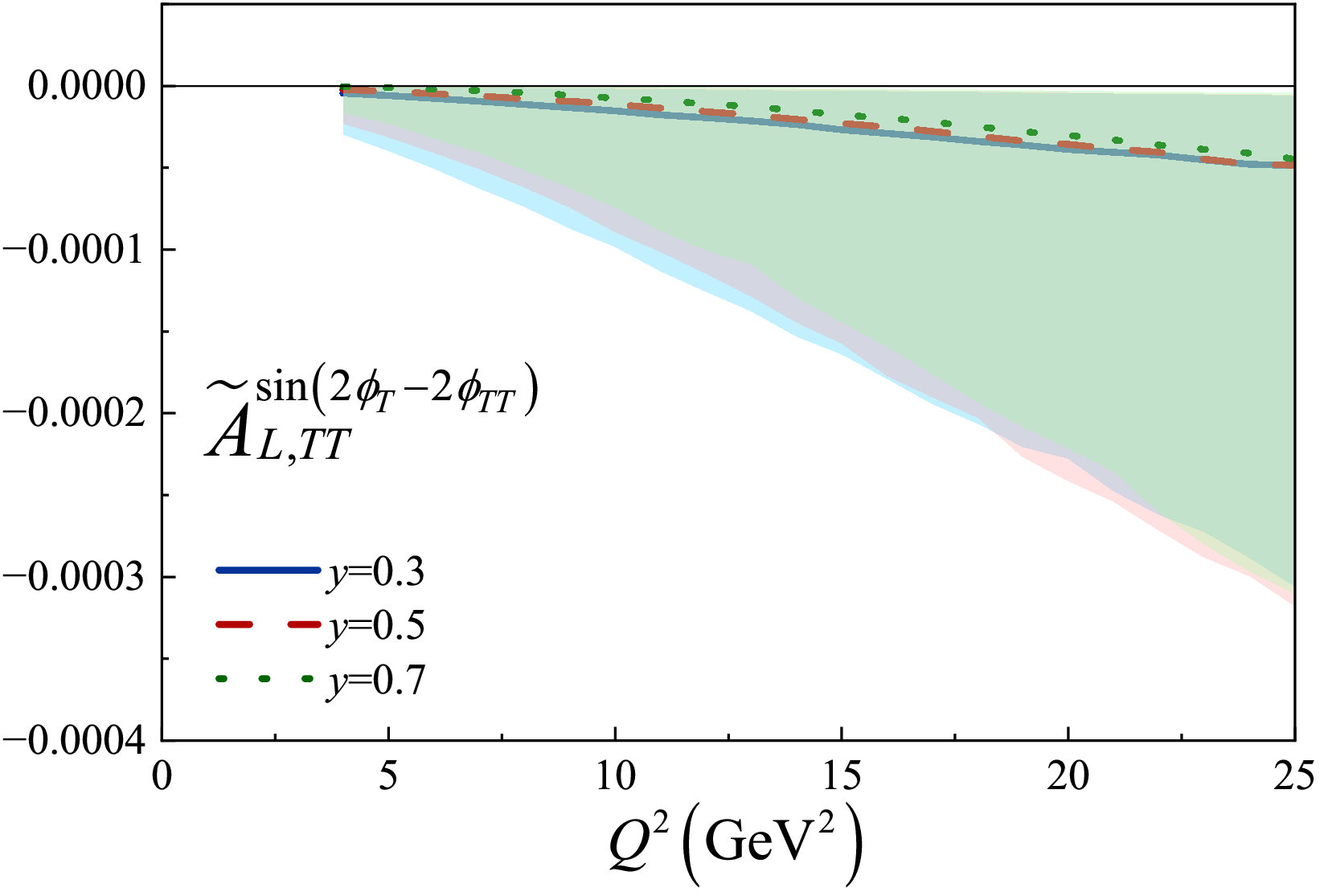}
\caption{Four normalized observables versus $Q^2$ at $q_T=0.3\,\mathrm{GeV}$ and $s_{eD}=4050\,\mathrm{GeV}^2$, in the same order as Fig.~\ref{fig:y}. Blue solid, red dashed, and green dotted curves denote $y=0.3$, $0.5$, and $0.7$. The two moments proportional to $B_0$ vanish at $Q^2=M_\psi^2=9.61\,\mathrm{GeV}^2$ for the adopted SV LDMEs. Bands show the spectator-model replica variation.}
\label{fig:q2}
\end{figure*}

Figure~\ref{fig:q2} illustrates the different origins of the zeros in the $Q^2$ dependence. For $\widetilde A_{U,LL}$, smaller $y$ corresponds to larger $x_g$ at fixed $Q^2$ and shifts the zero to lower virtuality. If $f_{1LL}^g(x_0,q_T^2)=0$, the corresponding value is $Q^2=ys_{eD}x_0-M_\psi^2$. This shift is due to the momentum-fraction dependence of the model TMD, not to TMD evolution.

The two moments involving $h_{1LT}^g$ and $h_{1TT}^g$ are negative below $Q^2=M_\psi^2$, vanish there for all three values of $y$, and become positive above it. This common zero follows from $\mathcal B=0$ for the SV LDMEs and holds for every model replica. Consequently, the uncertainty bands also vanish at that point; this does not imply that the gluon TMDs are precisely known. Other LDME choices would shift the zero according to Eq.~\eqref{eq:zero-general}.

By contrast, the $g_{1TT}^g$ sine moment has no zero enforced at $Q^2=M_\psi^2$, because it is proportional to $C_0$ rather than $B_0$. Its central values remain small compared with the cosine moments, while the replica bands show substantial model-parameter dependence. This difference provides a way to distinguish the zero of a hard-scattering coefficient from zeros of the TMDs.

\begin{figure*}[t]
\centering
\includegraphics[width=0.485\textwidth]{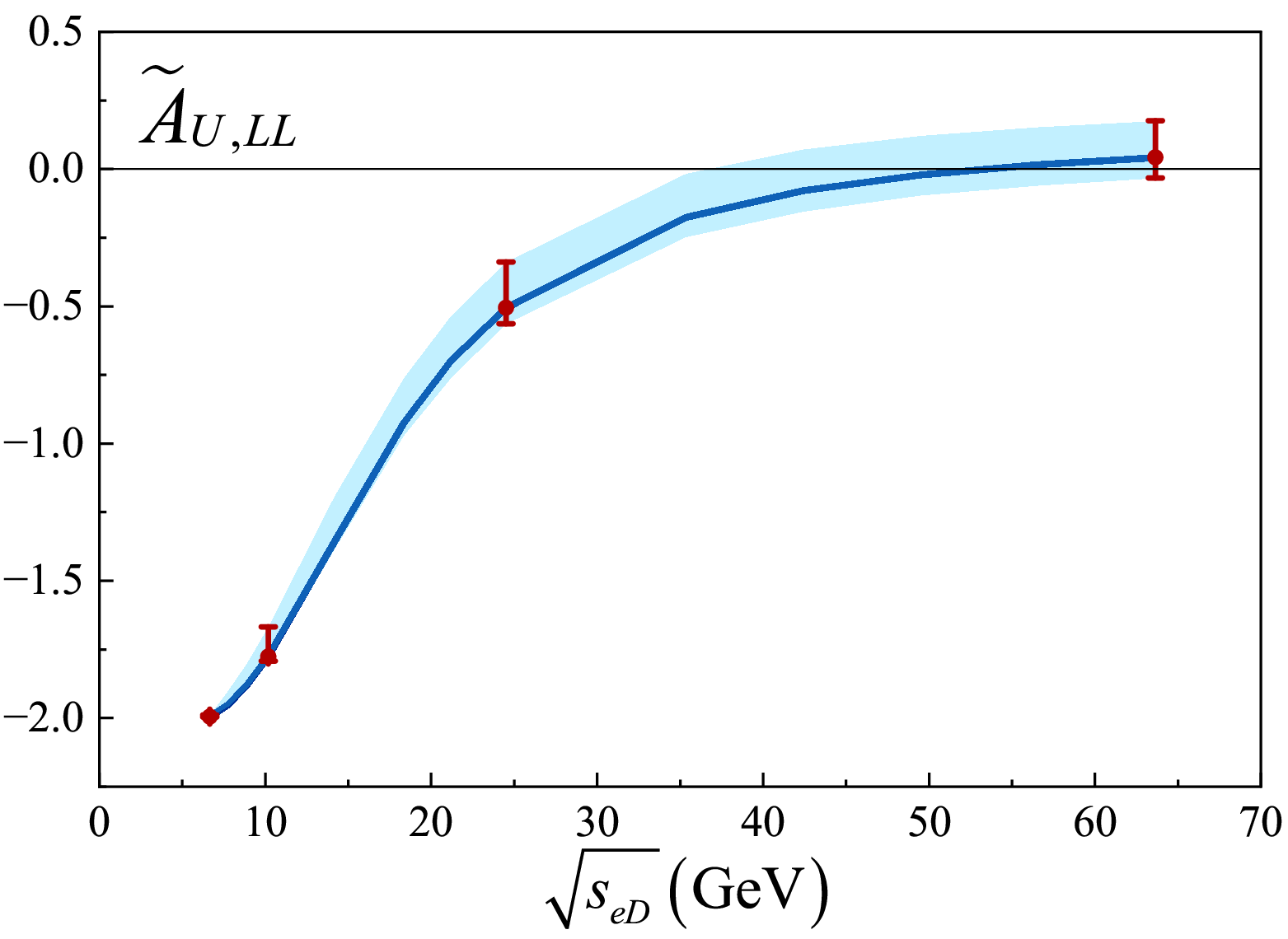}\hfill
\includegraphics[width=0.485\textwidth]{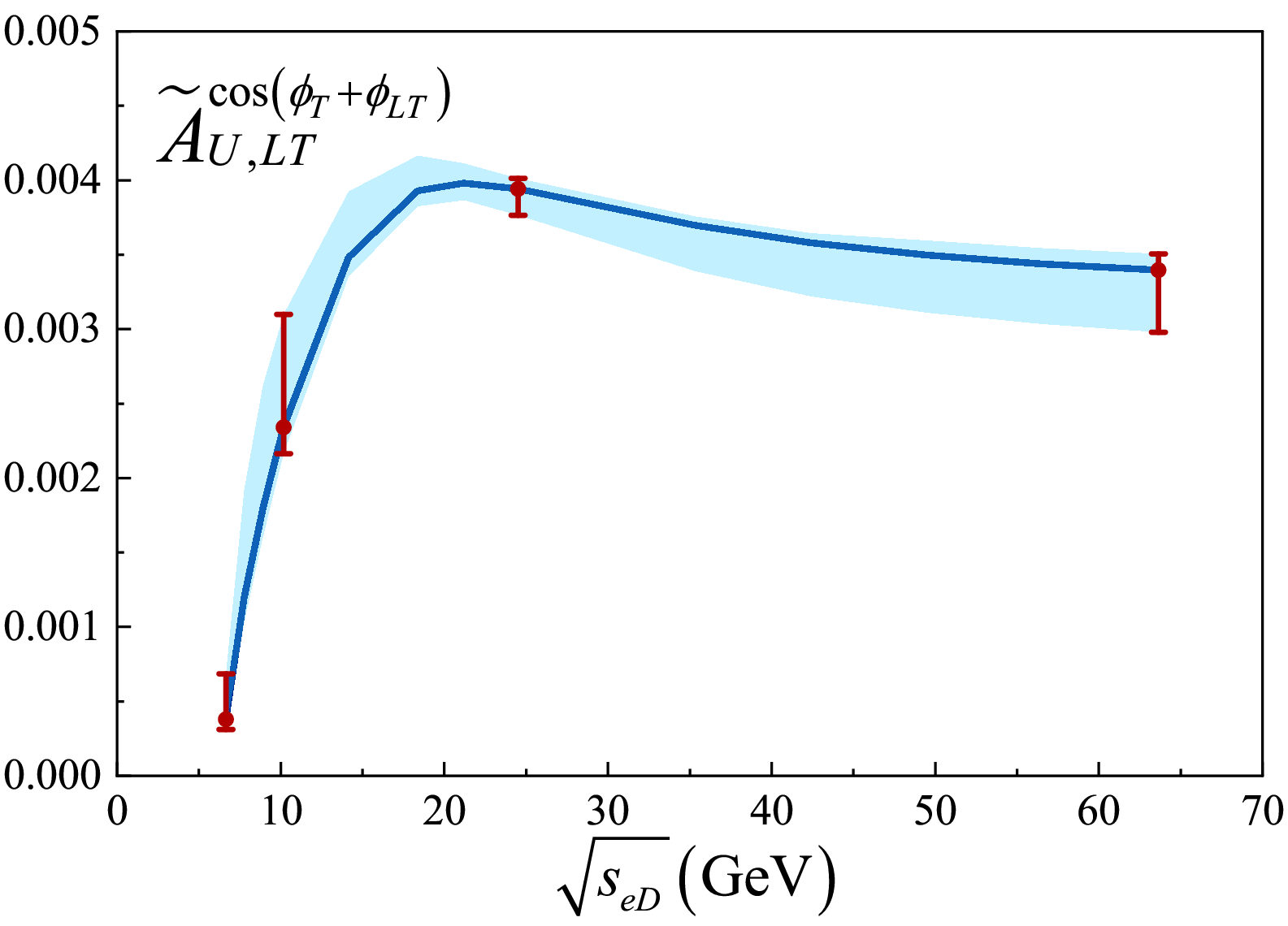}\\[5pt]
\includegraphics[width=0.485\textwidth]{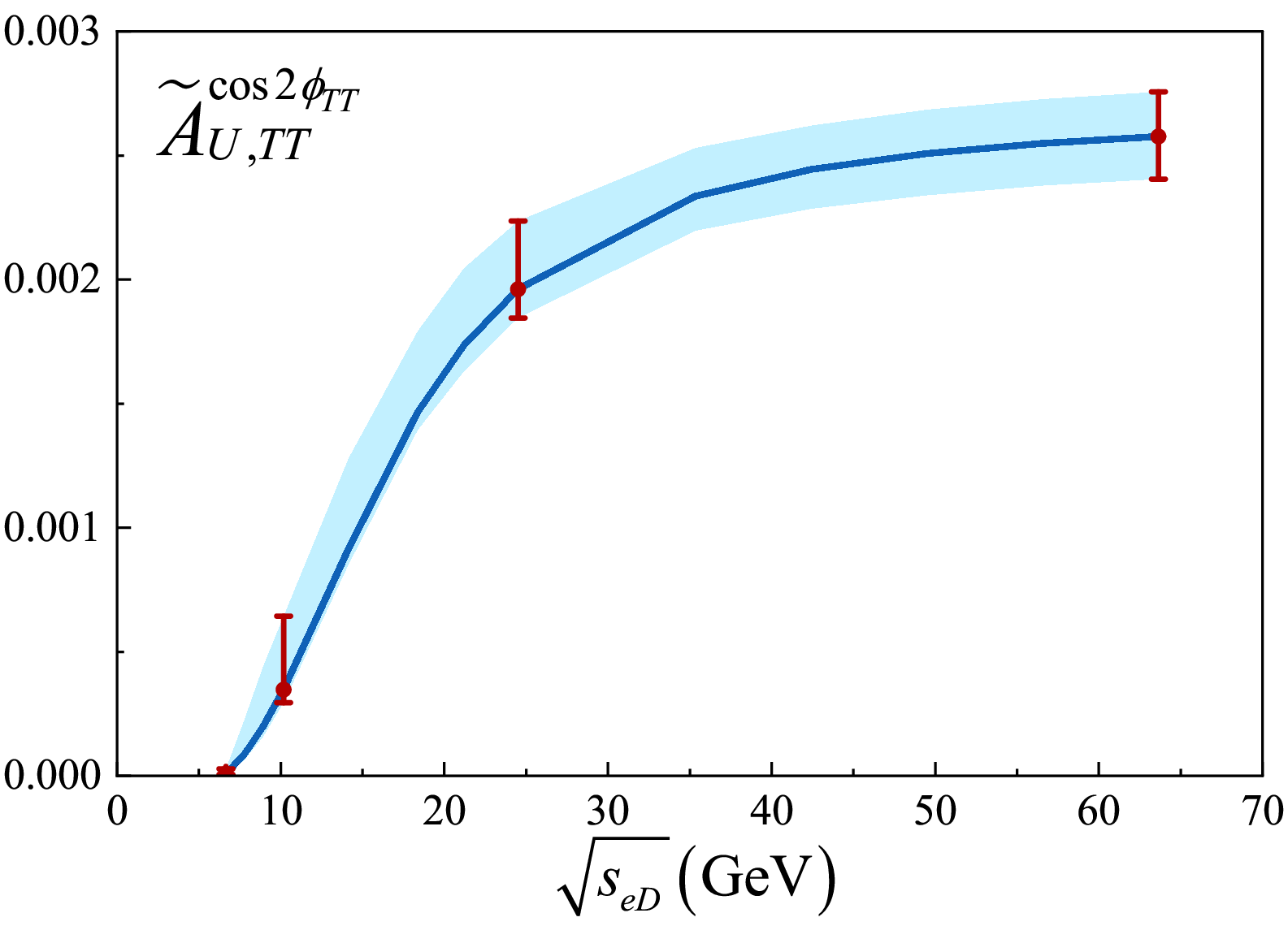}\hfill
\includegraphics[width=0.485\textwidth]{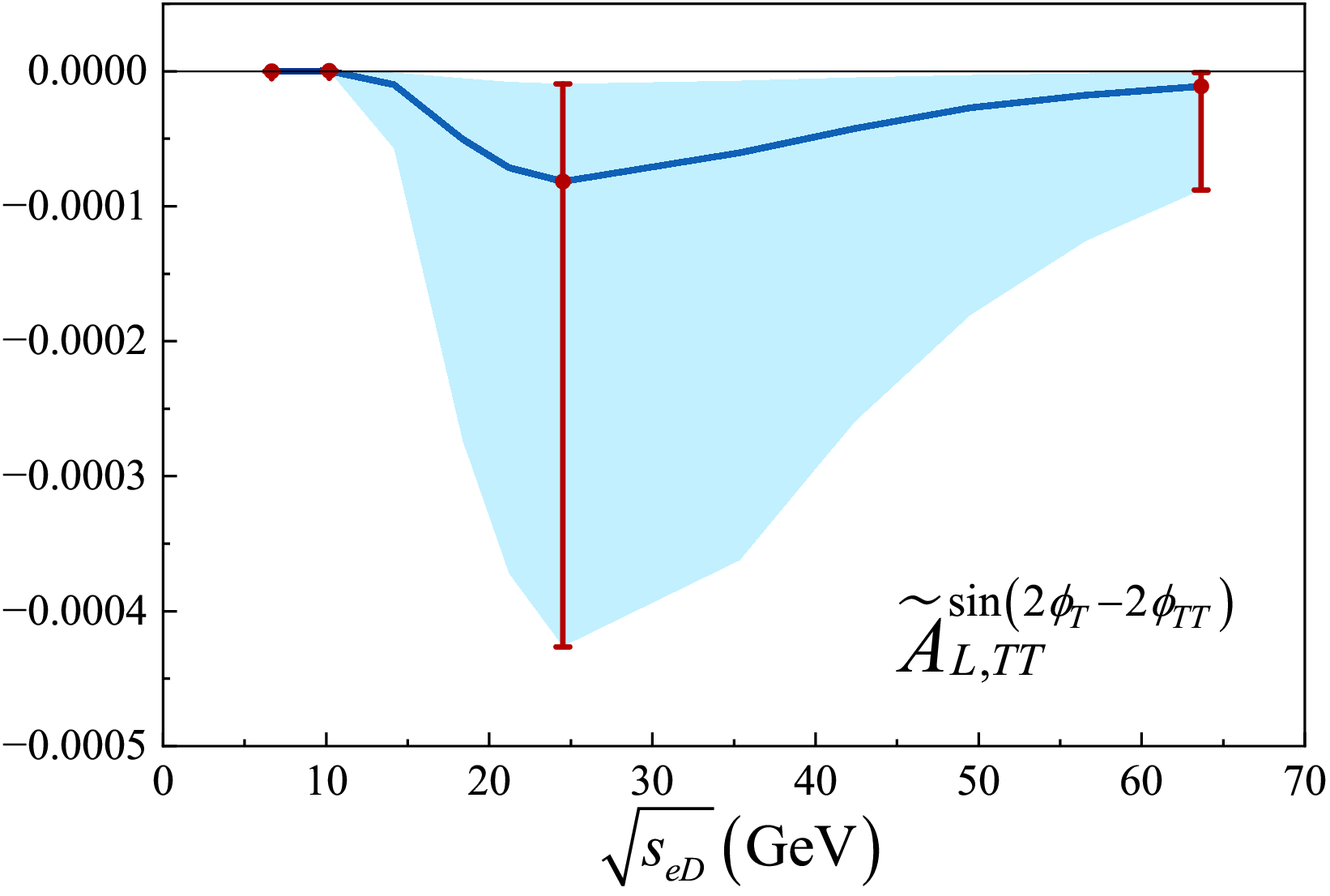}
\caption{Energy dependence of the four normalized observables, in the order of Fig.~\ref{fig:y}, at $Q^2=10\,\mathrm{GeV}^2$, $y=0.5$, and $q_T=0.3\,\mathrm{GeV}$. Blue curves join the calculated replica-65 values, and cyan bands show the replica intervals. Red markers and bars indicate $\sqrt{s_{eD}}\simeq6.65$, $10.18$, $24.51$, and $63.64\,\mathrm{GeV}$ and their model intervals; they are not experimental data. The lowest-energy point and the curve below $\sqrt{s_{eD}}\simeq8.12\,\mathrm{GeV}$ are below the hadronic production threshold and are shown only as formal evaluations. The second reference point, at $x_g=0.3783$, lies in the large-$x_g$ region where the spectator-model estimates are less constrained.}
\label{fig:s}
\end{figure*}

Figure~\ref{fig:s} shows the energy dependence of the same four observables at the fixed setting in Eq.~\eqref{eq:s-scan}. In this scan,
\begin{align}
x_g=\frac{39.22\,\mathrm{GeV}^2}{s_{eD}},
\label{eq:xg-energy}
\end{align}
while $B_0/A_0$, $C_0/A_0$, and the explicit factors of $q_T/M_D$ are unchanged. The energy scan thus probes the $x_g$ dependence of the TMD ratios at fixed $Q^2$, $y$, and $q_T$, without changing the ratios of hard-scattering coefficients.

The $LL$ ratio increases with $\sqrt{s_{eD}}$ and crosses zero around $50$--$55\,\mathrm{GeV}$, corresponding to the same $x_g$-dependent sign change seen in Figs.~\ref{fig:y} and \ref{fig:q2}. Its large negative values at the lowest energies are formal model evaluations and should not be interpreted as physical asymmetries below the hadronic production threshold.

The $h_{1LT}^g$ and $h_{1TT}^g$ moments have different energy dependences although both contain the same ratio $B_0/A_0$. The former reaches a broad maximum near $\sqrt{s_{eD}}=20\,\mathrm{GeV}$, while the latter increases more gradually toward the highest energies. Since their hard coefficients are unchanged in this scan, the difference follows from the $x_g$ dependence of $h_{1LT}^g/f_1^g$ and $h_{1TT}^g/f_1^g$.

The $g_{1TT}^g$ sine moment reaches its most negative central values around $\sqrt{s_{eD}}=20$--$25\,\mathrm{GeV}$, with a value of approximately $-8\times10^{-5}$ at the $24.51\,\mathrm{GeV}$ reference energy. Its replica interval is considerably wider than the central value, particularly in this intermediate-energy region. The distinct energy dependence of this T-odd moment reflects the $x_g$ dependence of $g_{1TT}^g/f_1^g$ in the spectator model.

The low-energy part of Fig.~\ref{fig:s} also tests the limits of the kinematic approximation. Table~\ref{tab:energies} shows that the first two reference energies probe large deuteron momentum fractions. More decisively, retaining the target mass gives the hadronic invariant mass
\begin{align}
W_D^2=(P+q)^2=M_D^2+y(s_{eD}-M_D^2)-Q^2.
\label{eq:wd-threshold}
\end{align}
Baryon-number conservation requires a baryon-number-two remnant in addition to the $J/\psi$. A necessary condition is therefore $W_D\geq M_\psi+M_D$, which gives $\sqrt{s_{eD}}\gtrsim8.12\,\mathrm{GeV}$ for Eq.~\eqref{eq:s-scan}. At the lowest reference energy, $W_D\simeq3.72\,\mathrm{GeV}$ is smaller than $M_\psi+M_D\simeq4.98\,\mathrm{GeV}$, even though the approximate expression for $x_g$ gives a value below unity. The results below threshold are retained in Fig.~\ref{fig:s} solely to display the formal continuation of the model expressions, and they have no interpretation as physical production asymmetries. A separate limitation comes from the gluon momentum fraction: the second reference energy corresponds to $x_g=0.3783$, where the spectator-model predictions are sensitive to their large-$x_g$ behavior. If one restricts the comparison to $x_g\lesssim0.2$, the kinematics of Eq.~\eqref{eq:s-scan} require $\sqrt{s_{eD}}\gtrsim14\,\mathrm{GeV}$. This latter value is a conservative choice for discussing the model input, not a production threshold. Hence, the low-energy part of the scan should not be regarded as a quantitative forecast for the corresponding facilities.

The results show three qualitatively different sources of $Q^2$ and energy dependence. The observables in Eqs.~\eqref{eq:all0}, \eqref{eq:alt-minus}, and \eqref{eq:att-minus} depend only on gluon TMD ratios. The cosine moments proportional to $B_0$ additionally depend on the relative color-octet LDMEs and vanish where $\mathcal B=0$. The sine moments involving $g_{1LT}^g$ and $g_{1TT}^g$ instead depend on $C_0$ and display larger relative model uncertainties.

A comparison with experimental data requires the actual beam and target polarizations, detector acceptance, and event yields. In finite kinematic bins, the measured moments must be formed from ratios of appropriately weighted cross-section integrals, rather than by averaging the pointwise ratios plotted here. Background contributions and quarkonium-production kinematics must also be considered.

Finally, $\bm q_T=\bm k_T$ assumes negligible transverse recoil from soft radiation and the formation of the quarkonium state. More complete treatments include convolutions with quarkonium TMD shape functions and the associated scale dependence~\cite{Echevarria:2019ynx,Fleming:2019pzj,Boer:2023zit,Echevarria:2024idp,Echevarria:2025oab}. The present calculation does not include these effects, which may change the $q_T$ distributions and the predicted asymmetries.

\section{Conclusion}
\label{sec:summary}

We have studied the tensor-polarized gluon TMD contributions to low-$q_T$ $J/\psi$ electroproduction off the deuteron in leading-order photon-gluon fusion. We also display the unpolarized and vector-polarized cross sections, including the nonzero beam-helicity-dependent $g_{1T}^g$ contribution. Using the $LL$, $LT$, and $TT$ components of the spin-1 gluon correlator, we derived the corresponding cross-section terms and defined ten normalized observables, consisting of one azimuthally independent ratio and nine azimuthal moments. The beam-helicity-dependent sine moments provide access to the T-odd distributions $g_{1LT}^g$ and $g_{1TT}^g$.

Our spectator-model estimates show that the observables involving $f_{1LL}^g$, $f_{1LT}^g+h_{1LT}^g$, and $f_{1TT}^g-h_{1TT}^{\perp g}$ are determined by gluon TMD ratios, with the common hard-scattering coefficient canceling. The linearly polarized gluon moments proportional to $B_0$ have a stronger dependence on $Q^2$ and share a zero at $Q^2=M_\psi^2$ for the adopted SV LDMEs. This zero is fixed by the relative color-octet LDMEs, not by the gluon distributions. The T-odd moments are smaller in the present model and have substantial uncertainty bands. Their magnitude is also sensitive to the form factor $g_2$ entering the deuteron-gluon-spectator vertex.

The collision-energy scan probes the $x_g$ dependence of the model TMD ratios at fixed hard-scattering kinematics. Its low-energy extension also illustrates the importance of the hadronic production threshold and the limitations of the spectator model at large $x_g$. Our results are fixed-scale estimates based on a leading-order color-octet description with negligible quarkonium-formation recoil. A quantitative comparison with experiment will require TMD evolution, quarkonium TMD shape functions, and realistic treatments of polarization, acceptance, and backgrounds.

\section*{Acknowledgements}
This work is supported in part by the National Natural Science Foundation of China under Grant No. 12150013. Xiupeng Xie is also supported by the SEU Innovation Capability Enhancement Plan for Doctoral Students under Grant No. CXJH$\_$SEU 25138.

\appendix
\section{Hard-scattering coefficients before applying heavy-quark spin symmetry}
\label{app:hard}

For reference, define $\mathcal O_J=\langle0|\mathcal O_8^{J/\psi}({}^3P_J)|0\rangle$. The channel sums used in the calculation can be written as~\cite{Bacchetta:2018ivt,liu:2023unl}
\begin{align}
\mathcal A_1={}&\mathcal O_S+\frac{4}{3M_\psi^2(M_\psi^2+Q^2)^2}
\Big[(3M_\psi^2+Q^2)^2\mathcal O_0\notag\\
&+2Q^2(2M_\psi^2+Q^2)\mathcal O_1\notag\\
&+\frac25(6M_\psi^4+6M_\psi^2Q^2+Q^4)\mathcal O_2\Big],\label{eq:general-a1}\\
\mathcal A_2={}&\frac{16Q^2}{3(M_\psi^2+Q^2)^2}
\left(\mathcal O_1+\frac35\mathcal O_2\right),\label{eq:general-a2}\\
\mathcal B={}&-\mathcal O_S+\frac{4}{3M_\psi^2(M_\psi^2+Q^2)^2}
\Big[(3M_\psi^2+Q^2)^2\mathcal O_0\notag\\
&-2Q^4\mathcal O_1+\frac25Q^4\mathcal O_2\Big],\label{eq:general-b}\\
\mathcal C={}&\mathcal O_S+\frac{4}{3M_\psi^2(M_\psi^2+Q^2)^2}
\Big[(3M_\psi^2+Q^2)^2\mathcal O_0\notag\\
&+2Q^4\mathcal O_1+\frac25(-6M_\psi^4+Q^4)\mathcal O_2\Big].
\label{eq:general-c}
\end{align}
Applying $\mathcal O_1=3\mathcal O_P$ and $\mathcal O_2=5\mathcal O_P$ gives Eqs.~\eqref{eq:a1}--\eqref{eq:c}. The factors $1/3$ and $1/5$ associated with the spin-averaged $P$-wave projectors are thereby accounted for before using the multiplicity relation. This distinction avoids counting the $J$ multiplicities twice.

The harmonic expressions also follow directly from transverse tensor contraction. In two dimensions,
\begin{align}
S_{TT\alpha}^{\{i}k_T^{j\}\alpha}
=g_T^{ij}\,k_T^{\alpha\beta}S_{TT\alpha\beta}.
\label{eq:tt-trace}
\end{align}
This identity explains why $h_{1TT}^{\perp g}$ accompanies the same hard coefficient and the same harmonic as $f_{1TT}^g$, with the opposite sign in Eq.~\eqref{eq:tt-u}. It also shows why their separate extraction is not possible from the present channel. The nonzero trace of $S_{LT}^{\{i}k_T^{j\}}$ analogously produces the shared $LT$ cosine response in Eq.~\eqref{eq:lt-u}.

\end{document}